\documentclass[10pt]{article}
\usepackage{fancyhdr}
\usepackage{extramarks}
\usepackage{amsmath}
\usepackage{amsthm}
\usepackage{amsfonts}
\usepackage{siunitx}
\usepackage{tikz}
\usepackage[plain]{algorithm}
\usepackage{algpseudocode}
\usepackage{multirow}
\usepackage{booktabs}
\usepackage{graphicx}
\usepackage{subfigure}
\usepackage[colorlinks,linkcolor=black,anchorcolor=black,citecolor=black,urlcolor=blue]{hyperref}
\usepackage{amsmath,bm}
\usepackage{booktabs}
\usepackage{mathtools}
\usepackage{amssymb}
\usepackage{caption}
\usepackage{capt-of}
\usepackage{mciteplus}
\usepackage{cite}
\usepackage{mathrsfs}
\usepackage[title,titletoc,toc]{appendix}
\usepackage{xr}
\usepackage{parskip}
\usepackage{soul}
\usepackage{textcomp}
\usepackage[colaction]{multicol}
\usepackage[switch]{lineno}
\usepackage{lipsum}
\usepackage{etoolbox}
\usepackage{longtable}
\usepackage{array}
\usepackage{tablefootnote}
\usepackage{ragged2e}
\newcolumntype{C}[1]{>{\centering\arraybackslash}p{#1}}
\usetikzlibrary{automata,positioning}
\usepackage{adjustbox}
\usepackage{makecell}
\usepackage{pdflscape}
\usepackage{setspace}
\usepackage{color}
\DeclareUnicodeCharacter{E5F8}{}

\DeclareUnicodeCharacter{0301}{\'{e}}

\DeclareMathOperator{\im}{im}

\usetikzlibrary{automata,positioning}

\begin{document}

\title{TIDAL: Topology-Inferred Drug Addiction Learning}
\author{Zailiang Zhu$^{1}$, Bozheng Dou$^{2}$, Yukang Cao$^{1}$, Jian Jiang$^{2,4}$\footnote{
 		Corresponding author.		Email: jjiang@wtu.edu.cn }, Yueying Zhu$^{2}$, \\
		Dong Chen$^{4}$, Hongsong Feng$^{4}$, Jie Liu$^{2}$,  
Bengong Zhang$^{2}$, Tianshou Zhou$^{3}$, 
Guo-Wei Wei$^{4,5,6}$\footnote{
 		Corresponding author.		Email: weig@msu.edu} \\
$^{1}$School of Computer Science and Artificial Intelligence, Wuhan Textile University,\\ Wuhan, 430200, P R. China \\
$^{2}$Research Center of Nonlinear Science, School of Mathematical and Physical Sciences, \\Wuhan Textile University, Wuhan, 430200, P R. China \\
$^{3}$Key Laboratory of Computational Mathematics, Guangdong Province, and \\ School of Mathematics, Sun Yat-sen University, Guangzhou, 510006, PR China\\
$^{4}$Department of Mathematics, Michigan State University,\\ East Lansing, Michigan 48824, USA\\
$^{5}$Department of Electrical and Computer Engineering, Michigan State University,\\
 East Lansing, Michigan 48824, USA\\
$^{6}$Department of Biochemistry and Molecular Biology, Michigan State University,\\ 
East Lansing, Michigan 48824, USA}
\date{\today} 

\maketitle

\abstract{Drug addiction or drug overdose is a global public health crisis, and the design of anti-addiction drugs remains a major challenge due to intricate mechanisms. Since experimental drug screening and optimization are too time-consuming and expensive, there is urgent need to  develop  innovative artificial intelligence (AI) methods for  addressing the challenge. We tackle this challenge by topology-inferred drug addiction learning (TIDAL) built from integrating topological  Laplacian, deep bidirectional transformer, and ensemble-assisted neural networks (EANNs). The topological Laplacian is a novel algebraic topology tool that embeds molecular topological invariants and algebraic invariants into its harmonic spectra and non-harmonic spectra, respectively. These invariants complement sequence information extracted from  
a bidirectional  transformer. We validate the proposed TIDAL framework on  22 drug addiction related, 4 hERG, and 12 DAT datasets, showing that TIDAL is a state-of-the-art framework for the modeling and analysis of drug addiction data. We carry out cross-target analysis of the current drug addiction candidates to alert their side effects and identify their repurposing potentials, revealing drug-mediated linear and bilinear target correlations. Finally,  TIDAL is applied to shed light on relative efficacy, repurposing potential, and potential side effects  of  12 existing anti-addiction medications. Our results suggest that   TIDAL  provides a new computational strategy for pressingly-needed anti-substance addiction drug development.    }

 keywords: Persistent Laplacian;  Bidirectional transformer;  Drug addiction; side effect; repurposing; drug-mediated  bilinear target correlations; drug-mediated target-target networks.   

\maketitle

 {\setcounter{tocdepth}{4} \tableofcontents}
\newpage

\section{Introduction}

Drug addiction, clinically termed drug use disorder, is a fast-growing worldwide social and health problem with enormous economic and financial costs. 
For example, the United States (U.S.) recorded 70630 drug abuse-related deaths in 2019, and that number exceeded 100,000 in 2021, up 15 percent from 2020, becoming the leading cause of accidental deaths in the U.S. The current approaches to treating drug addiction can be classified into two categories, one directly targeting the drug receptor system and the other indirectly targeting non-drug receptor systems, like the dopamine and glutamate receptors  \cite{Cao2021}. For psychostimulant drugs, like cocaine, the main mechanism for their treatment is to inhibit dopamine re-uptake via blockading the dopamine transporter (DAT). When developing DAT inhibitors as medications, it is very important to avoid off-target binding causing dangerous side-effect, such as the blockade of a potassium channel, the human {\it ether-a-go-go} (hERG) which can lead to potentially lethal ventricular tachycardia and even death. For opioid treatment, opioid replacement therapy (ORT) involves replacing an opioid, such as heroin, with a longer-acting but less euphoric substance. Commonly used drugs for ORT are methadone or buprenorphine which are taken under medical supervision. As of 2018, buprenorphine/naloxone is preferentially recommended, as the addition of the opioid antagonist naloxone is believed to reduce the risk of abuse via injection or insufflation without causing impairment. At present, the clinical drugs used to treat opioid addiction, including the opioid receptor agonists methadone and buprenorphine, and the opioid receptor antagonist naltrexone, are limited by their abuse liability and poor compliance. Therefore, the development of effective medications with lower abuse liability and better potential for compliance is urgently needed. 

Currently, therapeutic development against drug addiction mainly targets various neural transporters. A major complication originates from the intricate molecular mechanisms of drug addiction, which involves synergistic interactions among proteins upstream and
downstream of neural transporters \cite{Gao2021}. Additionally, it is too time-consuming and expensive to test so many proteins in traditional in vivo or in vitro experiments. Moreover, experimental testing involving animals or humans usually raises serious ethical concerns. Therefore, for large-scale assays, various computer-aided or in silico approaches, especially, machine learning (ML) and deep learning (DL) technologies have become highly attractive for drug design and discovery. These approaches are very valuable in target identification,  candidate screening, and generating new druglike compounds for further consideration, \cite{gao2020generative}. ML technologies have also been applied to the lead optimization of different druggable properties, like partition coefficient, toxicity,  solubility, binding affinity, pharmacokinetics, repurposing existing drugs to new diseases, etc \cite{gao20202d}. Various technologies, such as graph convolutional networks (GCNs), convolutional neural networks (CNNs), recurrent neural networks (RNNs), generative adversarial networks (GANs), etc, have become popular for drug discovery and largely reduced the need for time-consuming and expensive experiments, and thus benefited human health and welfare. 
In the ML study of potential drugs,  molecules need to be represented by descriptors or features.  Self-supervised learning (SSL) is a relatively new natural language processing (NLP) algorithm \cite{li2017deep} and has been successfully applied to many different fields, such as image identify \cite{yuan2021tokens},  bioinformatics \cite{singh2021rna}, etc. SSL strategy can be utilized to pre-train an encoder model which can generate latent space vectors as molecular representations without 3D molecular structures \cite{devlin2018bert}.   For instance, bidirectional encoder representations from transformers (BERTs) was created  to pre-train deep bidirectional transformer representation from unlabeled texts \cite{devlin2018bert,chen2021algebraic}. This technology, designed for understanding sequential words and sentences in NLP, has been used for uncovering the basic constitutional mechanism of molecules represented by simplified molecular-input line-entry system (SMILES)  \cite{weininger1988smiles}. Unlabeled SMILES strings can be treated as text-based chemical sentences and be considered as inputs for SSL pre-training \cite{wang2019smiles}. At present, large public chemical databases like ZINC, PubChem, and ChEMBL make SSL an attractive option for molecular representation generation.

However, latent space representations from SSL are insensitive to stereochemical information, such as chirality \cite{bruice2014organic}, steric effects, cis-trans isomerism, and the dihedral angle \cite{blondel1996new}. Chirality is a symmetry property such that a chiral molecule cannot be superposed on its mirror image and has important effects on physical, chemical, and biological properties. Chirality molecules, like (R)-thalidomide and (S)-thalidomide, have totally different drug actions. Steric effects are caused by the coming close together of atoms or radical groups, which influence the shape and reactivity of ions or molecules and are critical to chemistry, biochemistry, and pharmacology. For instance, steric effects in enzyme reaction would enhance or reduce its catalytic activity. Cis-trans isomerisms often have different physical and chemical properties due to different spatial arrangements of atoms, like trans-1,2-dichlorocyclohexane and cis-1,2-dichlorocyclohexane. All these steric effects are not considered in the latent space representation of transformers or autoencoders, since the related steric molecules have the same input strings. Additionally, latent space representations ignore specific physical and chemical information about task-specific properties. For instance, hydrogen bond interaction or van der Waals interaction can play a more important role than covalent interactions in drug binding properties \cite{chi2010toxic}.
Stereochemical information and associated  physical and chemical properties all depend on three-dimensional (3D) structures of molecules. Especially, macromolecules associated with  opioid or cocaine addiction have very complex  structures \cite{verma20103d}.  Molecular structural complexity and high dimensionality are central challenges in the design of efficient 3D representations. Recently, a series of 3D molecular representations have been proposed to meet these challenges, based on advanced mathematics, such as algebraic topology \cite{cang2017topologynet,meng2020weighted}, differential geometry \cite{nguyen2019dg}, and algebraic graph theory \cite{nguyen2019agl}.  

Traditional topology and/or homology contain little geometric information, which limits their practical applications. Persistent homology (PH) overcomes this limitation by a multiscale representation of data \cite{zomorodian2005computing,edelsbrunner2008persistent, mischaikow2013morse}. It incorporates geometric information in topological description and bridges the gap between geometry and topology. 
Although persistent homology has had much success in computational biology and chemistry \cite{cang2017topologynet,ciocanel2021topological}, it is insensitive to the homotopic shape evolution of data.
 
In this work, we introduce topology-inferred drug addiction learning (TIDAL) to combine the advantages of 3D    persistent Laplacian (PL) and deep bidirectional transformers for learning drug addiction data. PL, a topological Laplacian, is a brand new algebraic topology tool that not only fully recovers PH's topological invariants but also captures the homotopic shape evolution of data \cite{wang2020persistent}. PL is designed to effectively delineate the stereochemistry of 3D biomolecular structures and thus complements bidirectional transformers. Ensemble learning and deep learning are the two most popular algorithms, each having advantages for certain types of data. We introduce an ensemble-assisted neural network (EANN) algorithm to automatically combine their advantages and build a robust model for a variety of datasets.  The proposed TIDAL is applied to  22 drug addiction-related, 4 hERG, and 12 DAT datasets, giving rise to the best prediction.  We investigate potential drug candidates that inhibit drug addiction targets, the side effect from agents blocking unintended targets, and the drug repurposing potential. Additionally, we interrogate the  efficacy and potential side effects of  12 existing anti-addiction medications or candidates.  The proposed TIDAL  framework not only achieves the state-of-the-art in the modeling, analysis, and prediction  drug addiction datasets but also 
sheds light on the side effect, repurposing potential, relative efficacy,  and hERG blockage, drug-mediated linear and bilinear target correlations.  

\section{Results}

\subsection{ Overview of topology-inferred drug addiction learning (TIDAL) }

\begin{figure}[!tpb]
\centering
\includegraphics[width=11cm]{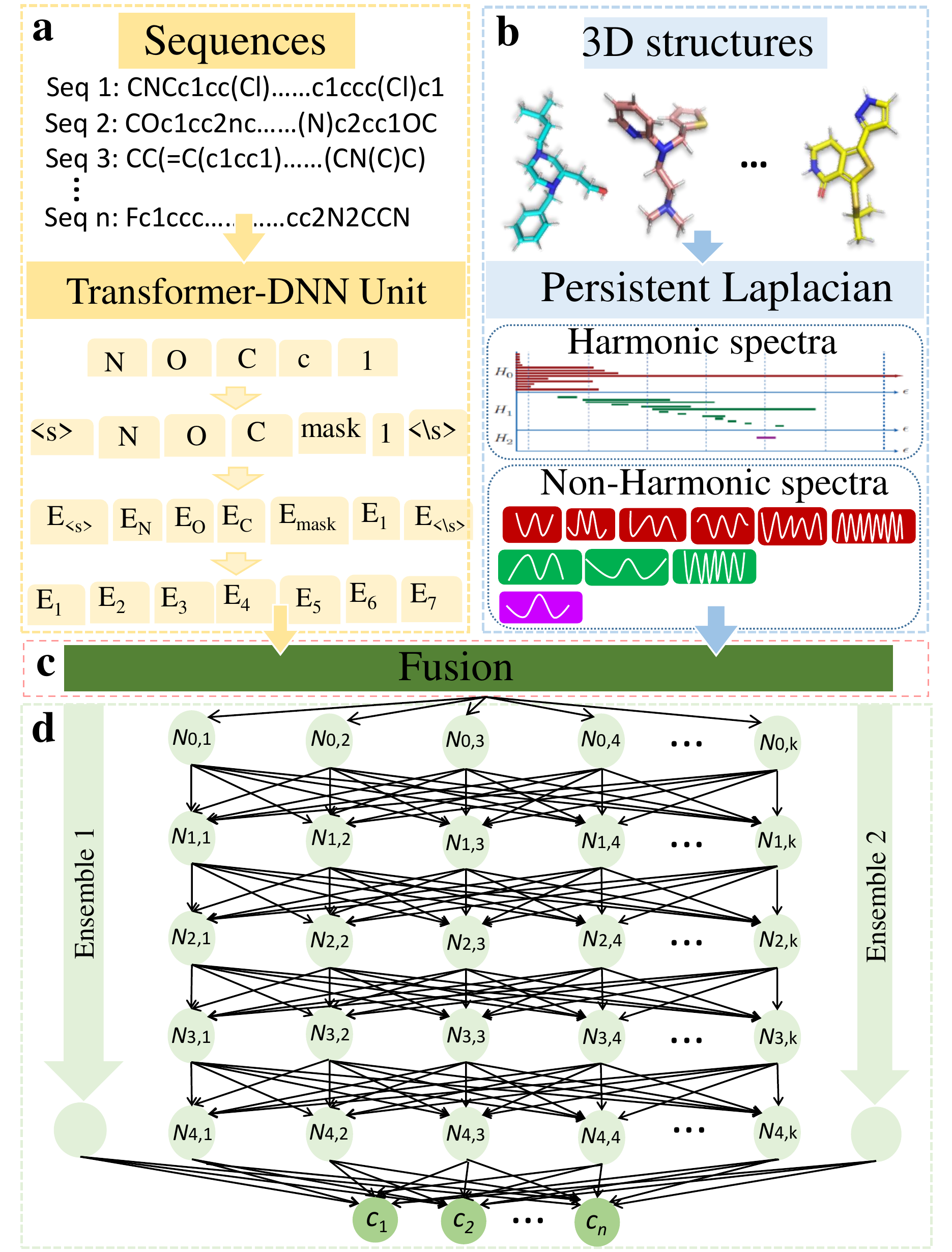}
\caption{
Illustration of the TIDAL platform for a classification task. {\bf a} Bidirectional transformer-based sequence embedding.  For a given molecule, its SMILES string is processed via a deep bidirectional transformer unit with pre-training or/and fine-tuning options to generate BT-FPs/$\rm{BT_f}$-FPs. {\bf b} Persistent Laplacian-based 3D structure embedding. The 3D structure of the given molecule is featured as PL-FPs by using persistent Laplacians to capture the topological persistence and geometric shape evolution through the calculations of harmonic and non-harmonic spectra, respectively. {\bf c} Concatenation of sequence and structure embeddings. {\bf d} Ensemble-assisted neural network (EANN) model. After embeddings, PL-FPs and BT-FPs (or $\rm{BT_f}$-FPs) are fed into ensemble algorithms to obtain regression predictions or classification probabilities. Meanwhile, PL-FPs and BT-FPs (or $\rm{BT_f}$-FPs) are employed for a deep neural network. At the last hidden layer of a deep neural network, semi-results (probabilities) from ensemble algorithms shown by two additional neural nodes on both sides of DNN are concatenated with other DNN nodes to achieve automatically weighted consensus between ensemble algorithms and DNN.  EANN automatically integrates DNN and ensemble algorithms and robustly improves the final prediction results for diverse datasets.}
\label{flowchart}
\end{figure}

 Figure   \ref{flowchart} shows the   TIDAL platform for classification, which consists of two complementary embedding modules, namely topology-based structural embedding and transformer-based sequence embedding, and a novel ensemble-assisted deep learning architecture. These embeddings are fed into deep neural networks and ensemble learning predictors. Predictions from deep learning and ensemble learning are automatically weighted to achieve optimal performance.  
We utilize hundreds of millions of molecular sequences available in various molecular databases to train our bidirectional transformer models as shown in Fig.   \ref{flowchart}{\bf a} \cite{chen2021algebraic}. The transformer is an SSL-based deep-learning model inspired by natural language processing. It learns synonyms and antonyms of a language for a better vocabulary and translates simple and short sentences from one  language to another one. Additionally, it utilizes an attention mechanism to single out certain words in the sentence to better understand the context. We use a deep bidirectional transformer (DBT) to learn basic constitutional rules from vast unlabeled SMILES data during an SSL-based pre-training process.  We denote the latent space vectors of DBT as bidirectional transformer-based fingerprints (BT-FPs). Note that, BT-FPs are insensitive to steric effects in many molecules.

Our idea is to complement transformer-based sequence embeddings with topology-based structural embeddings.  Fig.   \ref{flowchart}{\bf b} shows the topological embedding of 3D molecular structures.  Topology measures how the continuous deformation of a geometric object can cause a change in topological invariants. Persistent homology creates a family of multiscale geometric objects from filtration to enrich such a measurement, leading to the so-called topological persistence. However, neither topology nor persistent homology can capture the homotopic shape evolution of the data during the multiscale analysis. PL \cite{wang2020persistent}, a new topological Laplacian, overcomes this difficulty and offers a shape-aware representation of molecular structures. To further embed the physical and chemical interactions into the structural representations, we design element-specific PL descriptors based on statistic analysis of dataset compositions. The commonly occurring element types, such as C, H, O, N, etc. are grouped in pairs to describe various interactions, such as hydrogen bonds, electrostatics, hydrophilicity, and hydrophobicity. The resulting embeddings are called persistent Laplacian-based fingerprints (PL-FPs). 
	
In Fig.   \ref{flowchart}{\bf c}, complementary  BT-FPs and PL-FPs embeddings are fused to obtain a complete representation of underlying datasets.  The concatenated embeddings are fed into an ensemble-assisted neural network (EANN) model shown in  Fig.   \ref{flowchart}{\bf d}. EANN automatically weights ensemble learning and deep neural network models in a unified setting.  It is well known that deep neural network models are suitable for large datasets with nonlinear relationships between features and labels. They often outperform other machine learning methods in numerous chemical and biological applications. However, for small datasets,  deep neural network models might not match ensemble learning models, such as random forest (RF) and gradient-boosted decision trees (GBDT) \cite{jiang2020boosting}. RF and GBDT are the two most commonly used ensemble models that combine many weak learners into strong ones. Although both RF and GBDT use decision trees as weak learners, they are highly different methods. Essentially, RF needs no loss function and is simple to use, while GBDT utilizes a loss function to successively reduce errors and achieves better results. Both methods are suitable for small datasets which are very common for chemical and biological problems. To take advantage of ensemble models and deep neural network models, it is extremely common to construct a consensus in the literature. However, in such a consensus, results from different models are put on an equal footing, which may not be optimal for many problems.  In our EANN, we create an extended neuron layer to automatically balance the contributions from the deep neural network model and ensemble models.

\subsection{ Overview of persistent Laplacian } 

\begin{figure}[!tpb]
\centering
\includegraphics[width=10cm]{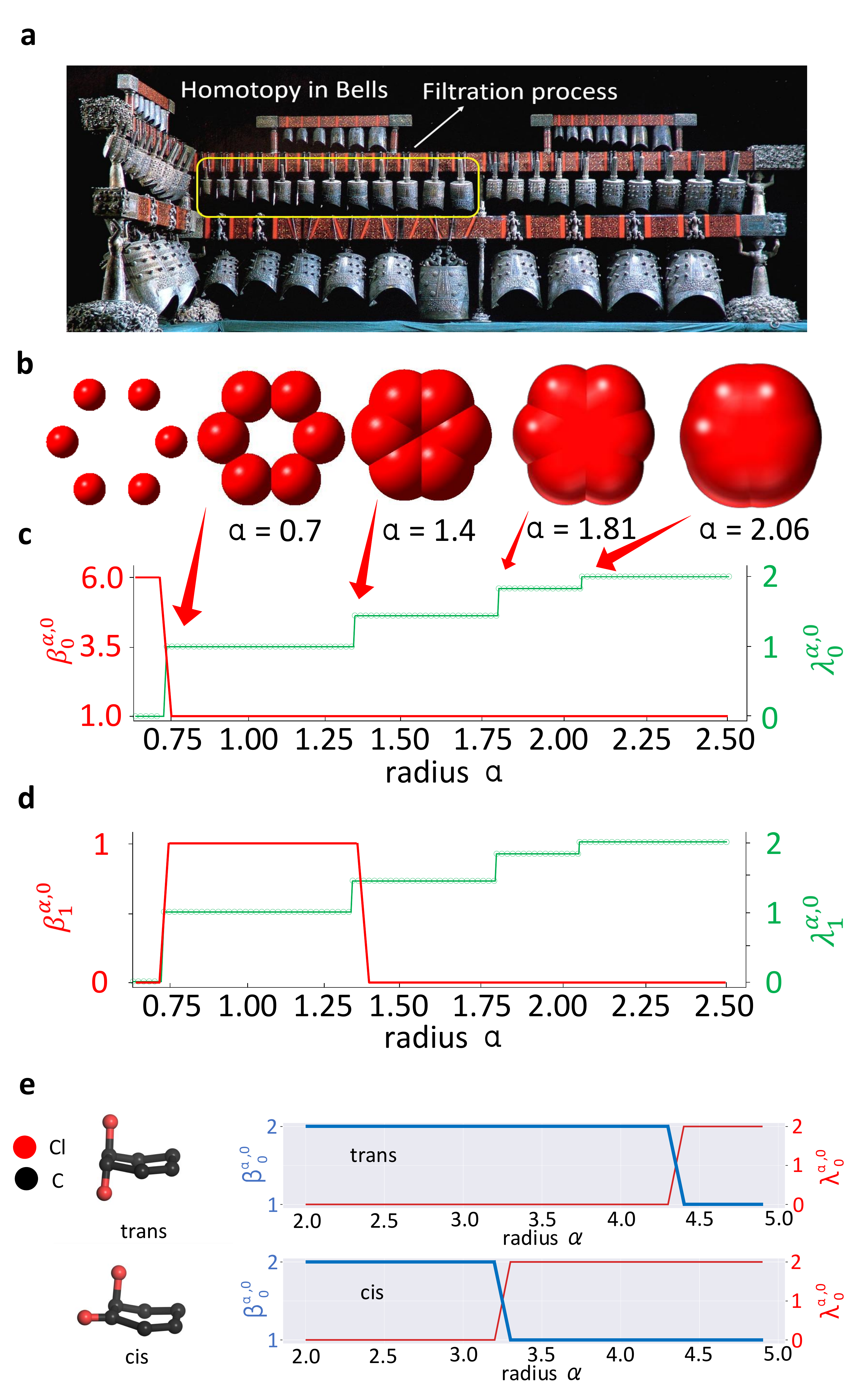}
\caption{Illustration of persistent Laplacian. 
{\bf a} Homotopy and filtration process in the Zenghouyi Chime Bells; {\bf b} The filtration process of C atoms of a benzene molecule with the increasing of atom radius. 
{\bf c} and {\bf d}  give the changes of persistent Betti numbers $\beta^{\alpha,0}_0$, $\beta^{\alpha,0}_1$, and the smallest nonzero eigenvalues of persistent Laplacian $\lambda^{\alpha,0}_0$, $\lambda^{\alpha,0}_1$ with the increasing of C atom radius $\alpha$ for the benzene molecule, respectively. The C atom radius is around 0.7 \AA.  
{\bf e}  The evolution of $\beta^{\alpha,0}_0$ and $\lambda^{\alpha,0}_0$ with the increasing of radius $\alpha$ for trans-1,2 (cis-1,2) dichlorocyclohexane molecule (right panel) and their different spatial 3D structure (left panel) with same SMILES string. }\label{bells}
\end{figure}

The persistent Laplacian is inspired by the wisdom of ancient  Zenghouyi Chime bells in China (see Fig.   \ref{bells}$\textbf{a}$). The   Zenghouyi Chime Bells, built in 433 B.C. in the Eastern Zhou Dynasty of ancient China have 65 bells in the whole set and are the largest in number, the best preservation, and the most complete rhythm found so far in the world.  These bells were designed, when struck, to produce distinct sounds with appropriate frequencies, covering a wide range of the spectrum. The physics behind the Chime bells is to create a nearly complete rhythm by a family of bells of varying sizes. The fundamental frequencies of chime bells systematically decrease as their sizes increase gradually. Since the fundamental frequency of a bell is its lowest frequency, it absorbs the highest amount of energy and is also perceived as the loudest when the bell is struck. The human ear identifies it as the specific pitch of the musical tone.  
 
From the perspective of topology, although these bells represent a deep mathematical filtration, they are homotopic to each other without any topological difference. Therefore, persistent homology cannot describe the shape evolution of these bells. We designed persistent Laplacians, aka persistent spectral graphs to capture homotopic shape evolution.  In particular, persistent Laplacian shares its foundation with persistent homology on the point cloud, simplex,  simplicial complex, and boundary operator.  However, harmonic spectra of   persistent Laplacian fully recover the topological persistence of persistent homology, while its non-harmonic spectra capture the shape evolution of  Chime Bells. Therefore,   our  persistent Laplacian can track the frequency changes of the Chime Bells.

 
We use the benzene molecule as another example to demonstrate the advantage of PL over PH. For simplicity, let us only consider the C atoms of benzene. Fig.   \ref{bells}$\textbf{b}$ shows the filtration process with the increasing radius $\alpha$ of the C atom. The behaviors 
 Betti-0 and  Betti-1 of PH  over filtration are illustrated as  the red lines Fig.   \ref{bells}$\textbf{c}$ and Fig.   \ref{bells}$\textbf{d}$, respectively. When atom radius is smaller than 0.7 \AA, there are six isolated C atoms and no ring exists, which corresponds to six counts in the Betti-0  ($\beta^{\alpha,0}_0 = 6$) and zero count in Betti-1 ($\beta^{\alpha,0}_1 = 0$).   When atom radius increases beyond 0.7 \AA, six C atoms form a ring, then the number of Betti-0 changes from six to one, and the number of Betti-1 becomes one. With the increasing of atom radius to 1.4 \AA, the ring is filled and the number of Betti-1 goes back to zero. When atom radius increases over 1.4 \AA, the six atoms undergo a homotopic evolution in which   Betti-0 and  Betti-1 do not change. However, as shown in  Fig.   \ref{bells}$\textbf{b}$,  the shape of the six carbon atoms evolves as the atom radius increases further, highlighting the inability and limitation of PH for molecular systems. 

In contrast, the lowest non-zero eigenvalues (  $\lambda^{\alpha,0}_0$  and   $\lambda^{\alpha,0}_1$)  of PL over the same filtration are illustrated by green circles in Fig.   \ref{bells}$\textbf{c}$ and Fig.   \ref{bells}$\textbf{d}$ for zero and 1st dimensional Laplacians, respectively.    PL captures not only all topological changes but also homotopic geometric evolution during filtration.   For instance, there are four jumps marked by four red arrows in Fig.   \ref{bells}$\textbf{c}$ corresponding to different connections between atoms under different atom radii in Fig.   \ref{bells}$\textbf{b}$, respectively. The value of the right $y$-axis is the smallest nonzero eigenvalues of PL and varies with different atom radii, which cannot be detected by the PH method.

Fig.   \ref{bells}$\textbf{e}$ illustrates the advantage of the PL method over transformers in dealing with molecules having steric effects. The cis-1,2 and trans-1,2 dichlorocyclohexane isomers are given in the left panel of Fig.   \ref{bells}$\textbf{e}$, omitting hydrogen atoms. These molecules have the same SMILES string ``C1CCC(C(C1)Cl)Cl", and thus the same transformer representation. However, the steric effects of these two molecules can be captured by the PL method as shown in the right panel of Fig.   \ref{bells}$\textbf{e}$ with only two chlorine atoms taken into consideration for simplicity. Both the harmonic spectra and 
non-harmonic spectra over filtration can distinguish these isomers.

\subsection{Overview of results}
\begin{figure}[!tpb]
\centering
\includegraphics[width=13cm]{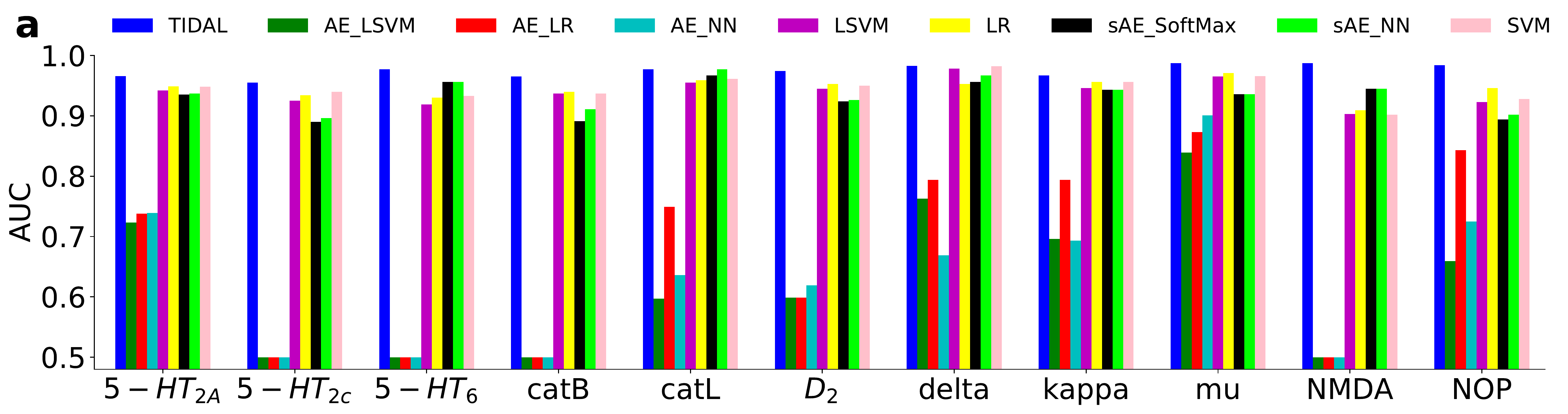}
\includegraphics[width=13cm]{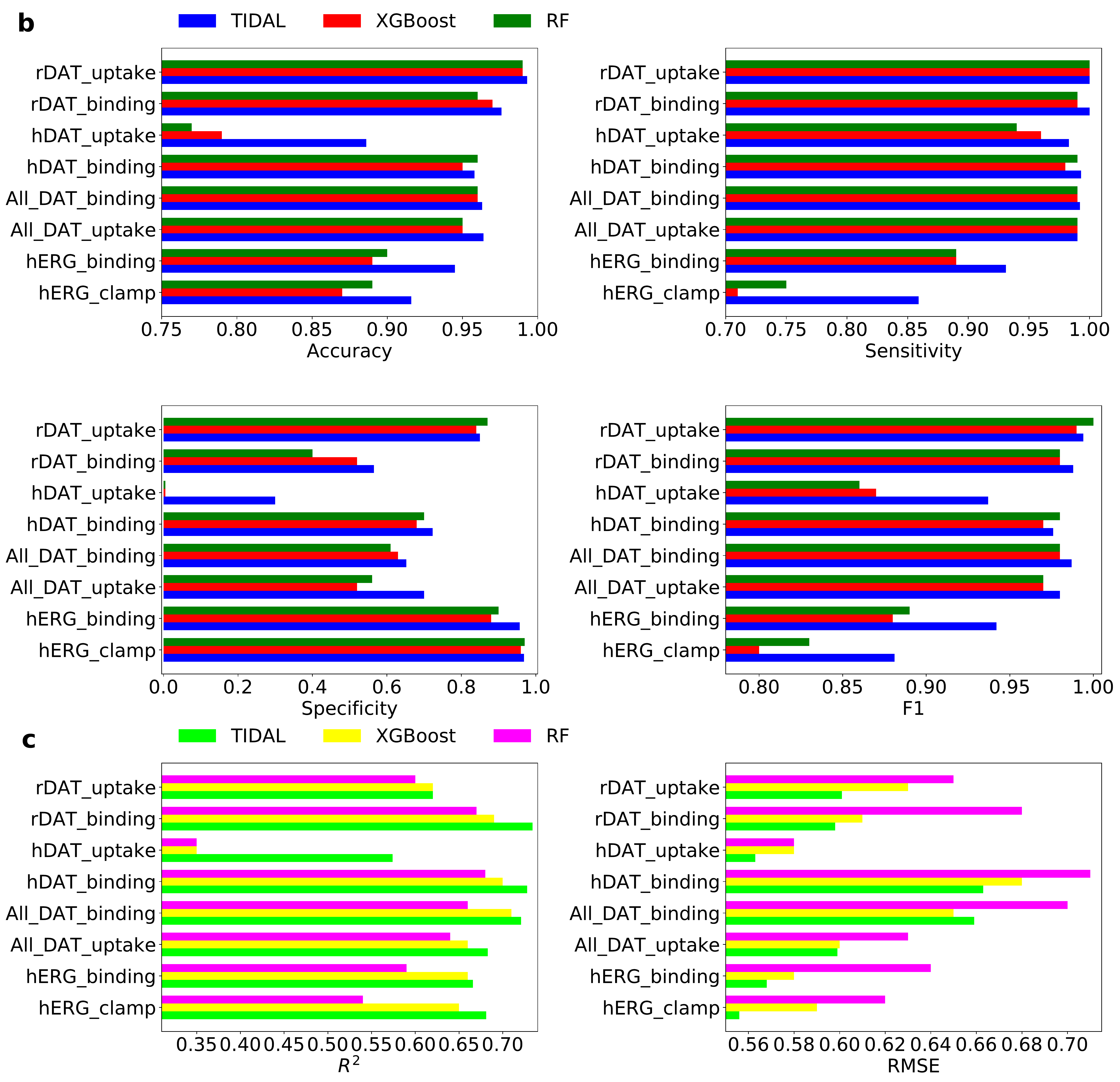}
\caption{Predicted results of classification and regression tasks on drug addiction-related, hERG, and DAT datasets. {\bf a} Comparison results of AUC for TIDAL model and other advanced models in \cite{warszycki2021pharmacoprint} on 11 drug addiction related classification datasets. TIDAL model outperforms these models including AE\_LR, LSVM, SVM, AE\_LSVM, AE\_NN, sAE\_NN, sAE\_SoftMax on 11 drug addiction related datasets. In addition, we also compare the results obtained by using our TIDAL model with those from other models in the literature \cite{czarnecki2015robust,smusz2015multi} on 5-HT$_{2A}$, 5-HT$_{2C}$, and 5-HT$_{6}$ datasets. Specifically, the values of AUC in our results are 8.3$\%$, 8.0$\%$, and 4.3$\%$ higher than those by Bayesian method \cite{czarnecki2015robust}, and are 33.8$\%$ higher than that obtained by homology model \cite{smusz2015multi} on 5-HT$_{6}$ dataset. {\bf b} gives the comparison results between our TIDAL model and other models (XGBoost and RF) in \cite{lee2021toward} on hERG and DAT classification datasets. The TIDAL model delivers the best accuracy, sensitivity, specificity, and F$_1$ score values on all these datasets compared with the XGBoost method \cite{lee2021toward}. When compared with the RF method \cite{lee2021toward}, the TIDAL model achieves better performance in 7/8, 8/8, 6/8, and 6/8 of datasets on the above four metrics, respectively. Additionally, on the all\_DAT\_binding, all\_DAT\_uptake, and hERG\_binding datasets, the TIDAL model yields higher $\rm{F_1}$ score compared with that by LV-FP, ECFP, Estate1, and Estate2 in \cite{Gao2021}, respectively. In terms of accuracy, the TIDAL model obtains almost the best or better results compared with that of LV-FPs and traditional 2D fingerprints \cite{Gao2021}. {\bf c} shows the comparison results between TIDAL model and XGBoost \cite{lee2021toward} and RF \cite{lee2021toward} on hERG and DAT regression datasets. About metrics coefficient of determination, $R^2$ and RMSE, the TIDAL model obtains both the best values than those by XGBoost \cite{lee2021toward} and RF \cite{lee2021toward} except on the all\_DAT\_binding dataset. We also compared the performance of the TIDAL model and that of methods in Ref.\cite{Feng2022} on the all\_DAT\_binding, all\_DAT\_uptake, and hERG\_binding datasets. For the metric of squared Pearson correlation coefficient ($P^2$), the TIDAL model performs better than or equivalently to that for different fingerprints in \cite{Feng2022}. Especially, RMSE values are raised up to 32$\%$, 26.9$\%$, and 33.8$\%$ on three datasets by the TIDAL model compared with the consensus of LV-FP and ECF8 \cite{Feng2022}, respectively.  }\label{comparison_cla_opioid_DAT_hERG}
\end{figure}

As shown in Fig. \ref{comparison_cla_opioid_DAT_hERG}, our TIDAL model gives a state-of-the-art performance on 22 drug addiction related, 4 hERG, and 12 DAT datasets for both classification and regression tasks in terms of a variety of evaluation metrics. 

\paragraph{Drug addiction related classifications dataset }

Fig. \ref{comparison_cla_opioid_DAT_hERG}$\textbf{a}$ shows the comparison of area under the  ROC (receiver operating characteristic) curve (AUC) between our TIDAL model and 8 other  state-of-the-art models \cite{warszycki2021pharmacoprint}  on 11 drug addiction related datasets for classification tasks. These models were created from pharmacophore fingerprints, which consist of 39973 bits and encode the presence, types, and relationships between pharmacophore features of a molecule and all detailed information about this molecular representation  \cite{warszycki2021pharmacoprint}. The dimensionality reduction algorithm was also used to reduce the length of molecular representations to 100 bits and enhance their performance. These  fingerprints were integrated with 8 ML algorithms \cite{warszycki2021pharmacoprint}, including 
logistic regression with autoencoder reduction algorithm (AE\_LR), 
linear support vector machines (LSVM), 
support vector machine (SVM), 
linear support vector machines with autoencoder reduction algorithm (AE\_LSVM), neural networks with autoencoder reduction algorithm (AE\_NN), 
neural networks with supervised autoencoder reduction algorithm (sAE\_NN), 
a single-layer linear classifier with supervised autoencoder reduction algorithm (sAE\_SoftMax), and 
logistic regression (LR). 
 TIDAL significantly outperforms these advanced ML, DL, and self-supervised learning models.
In particular, on the 5-HT$_{2C}$, 5-HT$_{6}$, and catB datasets,  AUC values obtained by TIDAL are  91.0$\%$, 95.4$\%$, and 93.0$\%$ higher than those of AE\_LSVM, AE\_LR, and AE\_NN \cite{warszycki2021pharmacoprint}, respectively. Note that, for the NMDA dataset with the smallest number of compounds 226, TIDAL's AUC is 97.4$\%$ higher than those achieved by AE\_LSVM, AE\_LR, and AE\_NN \cite{warszycki2021pharmacoprint}. These findings suggest that the proposed TIDAL model has many advantages over dimensionality-reduction-enhanced  ML, DL, and self-supervised learning models \cite{warszycki2021pharmacoprint} on small datasets. 	The details of values of AUC and Matthews correlation coefficient (MCC) on drug addiction-related classification datasets can be found in Supplementary Table 1. It worthy to note that Czarnecki et al. \cite{czarnecki2015robust} and Smusz \cite{smusz2015multi} also reported their studies of  5-HT$_{2A}$, 5-HT$_{2C}$, and 5-HT$_{6}$ datasets (See Table.\ref{table_1}). A detailed comparison with their results is given in the next section.

\paragraph{DAT and hERG dataset classifications}

TIDAL classification results and comparison on hERG and DAT datasets are illustrated in Fig. \ref{comparison_cla_opioid_DAT_hERG}$\textbf{b}$. Two models, namely, eXtreme Gradient Boosting (XGBoost) and random forest (RF) algorithms, were reported by   Lee et al. \cite{lee2021toward} The features employed in these models were selected from  200 different types of from \_descList of rdkit.Chem.Descriptors and the Gobbi 2D pharmacophore fingerprints. The 10   most correlated descriptors were utilized to optimize the performance of XGBoost and RF \cite{lee2021toward}. It is seen from Fig. \ref{comparison_cla_opioid_DAT_hERG}$\textbf{b}$ that: 
(1) TIDAL model attains better performance on all eight datasets than  the XGBoost method \cite{lee2021toward} on all evaluation metrics, i.e., 
accuracy, sensitivity, specificity, and $\rm{F_1}$ score. Particularly, on hERG\_clamp dataset, the sensitivity and $\rm{F_1}$ values of TIDAL are about   21.0$\%$ and 10.1$\%$ higher than those of the XGBoost method, respectively. In addition, TIDAL is 12.2$\%$ more accuracy than the   XGBoost method  on hDAT\_uptake dataset \cite{lee2021toward}; 
(2) For accuracy, the TIDAL model outperforms the RF method in 7/8 of datasets and is essentially as good as RF for the hDAT\_binding dataset \cite{lee2021toward}. On hDAT\_uptake dataset, the TIDAL model is  15.1$\%$ more accurate than the RF method \cite{lee2021toward};   
(3) in terms of metric sensitivity, the TIDAL model achieves the best values on all eight datasets compared with RF \cite{lee2021toward}. On hERG\_clamp dataset, the TIDAL model is 14.5$\%$ more sensitive than the RF model;  
(4) Our model achieves the best specificity in 6/8 of cases, except for hERG\_clamp dataset and  rDAT\_uptake dataset. For hDAT\_uptake dataset, both XGBoost and RF had nearly zero specificity. In contrast, the specificity of our TIDAL model is 0.3; and
(5) TIDAL model performs better than RF \cite{lee2021toward} in 6/8 of datasets on $\rm{F_1}$ score except for hDAT\_binding and rDAT\_uptake datasets. On hERG\_clamp dataset, the TIDAL model results in the growing up to 6.1$\%$ of $\rm{F_1}$ score. The details of four metrics values on hERG and DAT datasets can be found in Supplementary Table 2, and more metrics values of AUC and MCC are shown in Supplementary Table 3. Further comparison with Gao et al \cite{Gao2021} is given in the next section and the details of the comparison can be found in Table \ref{table_2_3}. 

\paragraph{DAT and hERG regression}

Regression analysis of hERG and DAT datasets is presented in Fig. \ref{comparison_cla_opioid_DAT_hERG}$\textbf{c}$. Comparison is given to TIDAL, XGBoost, and RF models \cite{lee2021toward} in terms of two metrics (coefficient of determination, $R^2$,  and RMSE). The TIDAL model outperforms XGBoost and RF in all cases except for the all-DAT binding dataset, where TIDAL and RF have a similar performance.  Specifically, our TIDAL model is 64.0$\%$ more accurate than   XGBoost, and RF models in terms of $R^2$ on the hDAT\_uptake dataset. 
Our TIDAL model is  12.7$\%$, 11.5$\%$, and 13.7$\%$ more accurate than RF in terms of  RMSE on  hERG\_binding, hERG\_clamp, and rDAT\_binding datasets,\cite{lee2021toward}  respectively. The details of $R^2$ and RMSE on hERG and DAT datasets are given in Supplementary Table 4 for TIDAL, XGBoost, and RF models. 
The predictive results in terms of another metric, the squared Pearson correlation coefficient ($P^2$), are given in Supplementary Table 5. 
More comparison with the results in the literature 	\cite{Feng2022} can be found in Table \ref{table_2_3} and is discussed in the next section. 
\paragraph{Results for 5-HT$_{2A}$, 5-HT$_{2C}$, and 5-HT$_{6}$ datasets }
\begin{table*}[h]\footnotesize
	\centering
     \caption{Comparison of classification results on 5-HT$_{2A}$, 5-HT$_{2C}$, and 5-HT$_{6}$ between TIDAL model and other methods \cite{czarnecki2015robust,smusz2015multi}.}\label{table_1}
    \begin{tabular}{cccccccc}
        \hline
        Dataset & \multicolumn{2}{c}{5-HT$_{2A}$} & \multicolumn{2}{c}{5-HT$_{2C}$} &\multicolumn{3}{c}{5-HT$_{6}$} \\
        \cmidrule(r){2-3} \cmidrule(r){4-5} \cmidrule(r){6-8}
        Model & Bayesian \cite{czarnecki2015robust} & TIDAL & Bayesian \cite{czarnecki2015robust} & TIDAL & Bayesian \cite{czarnecki2015robust} & Homology \cite{smusz2015multi} & TIDAL\\
        AUC & 0.892 & \textbf{0.966} & 0.884 & \textbf{0.955}&0.937 & 0.730 & \textbf{0.977}\\
        \hline
    \end{tabular}
\end{table*}

Having illustrated that TIDAL outperforms 8 other machine learning models in Fig.\ref{comparison_cla_opioid_DAT_hERG}$\textbf{a}$, we further carry out  additional   performance comparison  for 5-HT$_{2A}$, 5-HT$_{2C}$, and 5-HT$_{6}$ datasets (see	Table \ref{table_1}). Two new models, Bayesian \cite{czarnecki2015robust}  and homology \cite{smusz2015multi}, are included in this  AUC analysis.  Among them, the Bayesian model optimizes   Estate, extended, Klekota-Roth, MACCS, and Pubchem fingerprints, with the support vector machine algorithm \cite{czarnecki2015robust}. The homology model was constructed by integrating five machine learning algorithms, i.e., naive Bayes, sequential minimal optimization, $k$-nearest neighbor algorithm, decision tree, and random forest, with two features, namely structural interaction fingerprints and Spectrophores descriptors \cite{smusz2015multi}. As indicated in Table \ref{table_1}, our TIDAL model performs significantly better than these two models.

\paragraph{Results for all\_DAT\_binding,  all\_DAT\_uptake, and hERG\_binding datasets  }

\begin{table*}[htpb]\small
	\centering
     \caption{Comparison of classification and regression results on hERG and DAT datasets between TIDAL model and other methods \cite{Gao2021}.}\label{table_2_3}
    \begin{adjustbox}{center}
    \setlength{\tabcolsep}{4pt}{
    \begin{tabular}{ccccccccc}
        \hline
       Task& Dataset &Metric & TIDAL & \makecell{LV-FP\\ \cite{Gao2021} }& \makecell{ECFP\\ \cite{Gao2021}} & \makecell{Estate1\\ \cite{Gao2021}} & \makecell{Estate2\\ \cite{Gao2021}} & \makecell{Consensus of \\ LVFP + ECFP \\ \cite{Gao2021}} \\
       \hline
        \multirow{3}{*}{Classification}  &all\underline{ }DAT\underline{ }binding  & $\rm{F}_1$ & \textbf{0.987}& 0.980 & 0.980 & 0.980 & 0.980 & 0.980\\
        & all\underline{ }DAT\underline{ }uptake& $\rm{F}_1$ & \textbf{0.980}& 0.980&0.980&0.980&0.980&0.980\\
        & hERG\underline{ }binding&$\rm{F}_1$& \textbf{0.942}&0.880& 0.880 & 0.870 & 0.870 & 0.880\\
        \hline
        \multirow{3}{*}{Regression}&all\underline{ }DAT\underline{ }binding &  RMSE & \textbf{0.659} & 0.890&0.900&0.990&0.990&0.870\\
        & all\underline{ }DAT\underline{ }uptake & RMSE& \textbf{0.599}&0.800&0.780&0.860&0.890&0.760\\
        & hERG\underline{ }binding & RMSE& \textbf{0.568}& 0.800&0.800&0.870&0.870&0.760\\
        \hline
    \end{tabular}
    }
    \end{adjustbox}
\end{table*} 

To further analyze the performance of the proposed TIDAL model on 
all\_DAT\_binding,  all\_DAT\_uptake, and hERG\_binding datasets,  we consider 
four other models based on GBDT and four descriptors, namely latent-vector fingerprint (LV-FP),  ECFP, Estate1, and Estate2 \cite{Gao2021}. LV-FP is extracted from our transformer, whereas  ECFP, Estate1, and Estate2 are 2D fingerprints \cite{Gao2021}. The consensus of LVFP and ECFP was also presented. Table \ref{table_2_3} illustrates the $\rm{F_1}$ scores for various competing models. The TIDAL model displays similar or better classification than other methods. In particular, for the hERG\_binding dataset, our TIDAL model achieves  6.2$\%$ and 7.0$\%$ higher   $\rm{F_1}$ scores than LV-FP and ECFP do \cite{Gao2021}, respectively.  

 
Moreover, Gao et al.\cite{Gao2021}   have carried out regression studies on all\_DAT\_binding, all\_DAT\_uptake and hERG\_binding datasets \cite{Gao2021}.  They combined GBDT and  LV-FPs and 2D fingerprints, like ECFP, Estate1, Estate2, as well as the consensus of LV-FPs and ECFP. As shown in Table \ref{table_2_3}, our TIDAL model yields the RMSE values of   0.659, 0.599, and 0.568 on all\_DAT\_binding, all\_DAT\_uptake, and hERG\_binding datasets, respectively. 
These results are about 32$\%$, 26.9$\%$, and 33.8$\%$ better than the best result   in the literature for three datasets, respectively. 


\section{Discussion}
%
	
\subsection{Side effect and repurposing}

The mechanism of drug addiction is still not very clear, involving many proteins, like DAT, and D$_3$R for cocaine addiction and delta, kappa,   mu, and nociceptin opioid receptor (NOP) for opioid addiction  as well as hERG, a crucial potassium ion channel for potential side effects \cite{Gao2021}. These proteins form a protein-protein interaction (PPI) network, having a complex relationship. On the one hand, they are drug addiction treatment targets. On the other hand, they  may bring unexpected off-target effects for potential drugs.  Hence, it is necessary to systematically investigate potential drug candidates that inhibit   drug addiction targets, the  side effect from agents blocking unintended targets, and the drug repurposing potential \cite{Gao2021}. We carry out this exploration by cross-target binding affinity (BA) predictions. The basic idea is to systematically predict the BAs of one target's inhibitors with respect to another target. Specifically, a drug candidate with a high BA (the absolute value of BA) for its original target may also have   high BAs for other human proteins, indicating strong side effects, which can disqualify its candidacy for the original target.  Additionally, if a drug candidate binds weakly to its designated target but has a high BA with another unintended target, then it has a repurposing potential. Here, we build ML models for 13 proteins that have sufficient inhibitor data to build such models based on drug addiction related regression datasets. These models are used for drug side effect and repurposing analysis. 
Notably, our analysis reveals the drug-mediated linear and  bilinear target correlations and associated drug-mediated target networks.

\begin{figure}[!tpb]
\centering
\includegraphics[width=15cm]{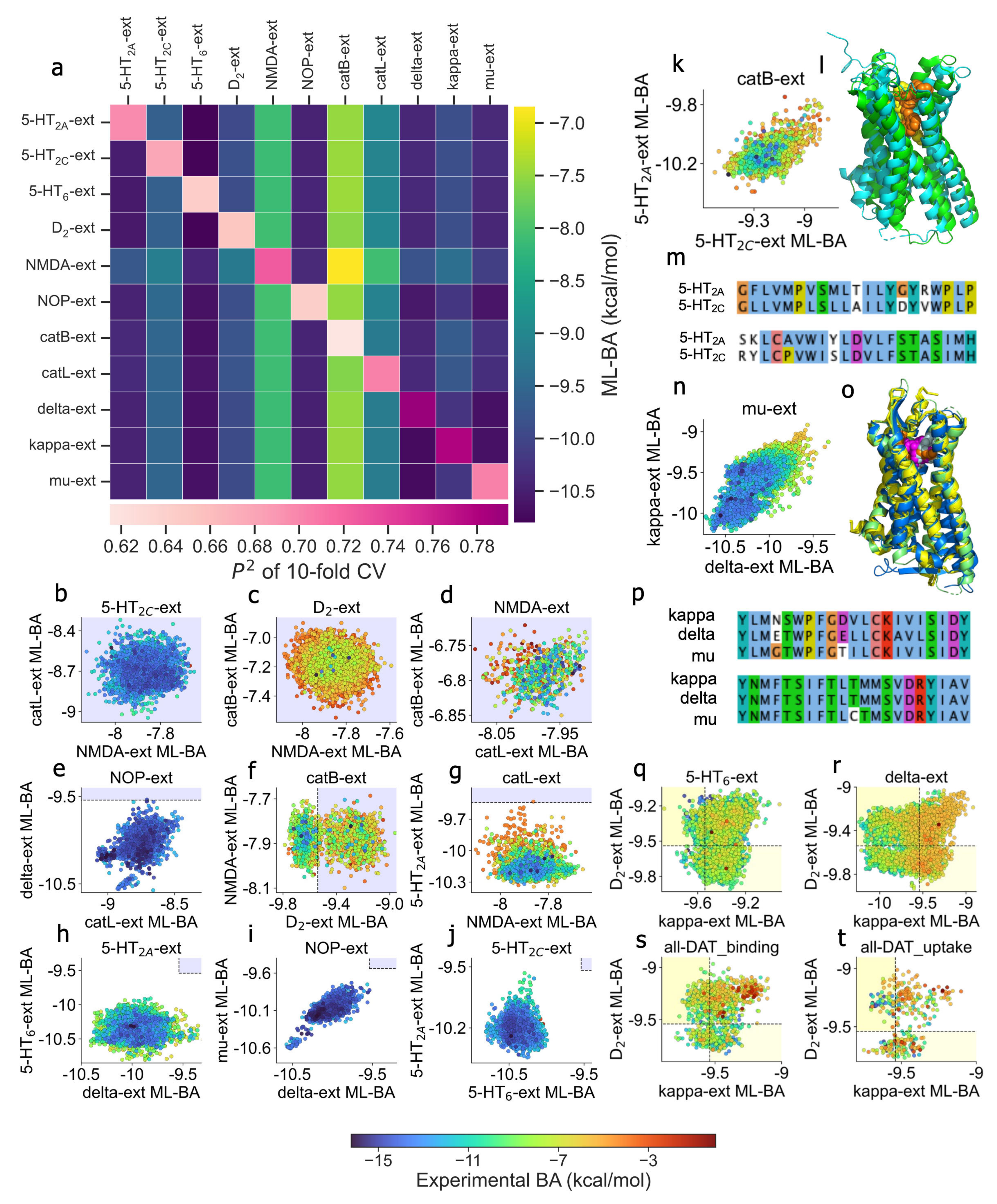}
\caption{Analysis of drug side effects and repurposing based on cross-target BA predictions. $\textbf{a}$ Heat map of cross-target BA predictions. The diagonal elements show the squared Pearson correlation coefficients of 10-fold cross-validation ($P^2$ of 10-fold CVs) on the machine-learning predicted BAs (ML-BAs) of targets. Other elements represent the highest ML-BAs among the inhibitors in each dataset to other targets.  $\textbf{b}$-$\textbf{j}$ Nine typical examples of cross-target BA predictions of potential drug side effects. In each ML-BA correlation plot, the title is the name of the dataset, and the colors of points indicate the experimental BAs for the designated target. The $x-$ and $y-$axes represent the predicted ML-BAs for two other proteins, respectively. The first, second, and third row show the cases with substantial side effects of potent inhibitors on zero, one, and two targets, respectively. The light blue background color outlines the optimal ranges without side effects on both targets (the values of BAs on $x-$ and $y-$axes are both larger than -9.54 kcal/mol). $\textbf{k}$ Illustration of  linear correlations in cross-target ML-BAs between 5-HT$_{2A}$ and 5-HT$_{2C}$, which reveals their binding site similarity and network connectivity. $\textbf{l}$ 3D structure alignment of 5-HT$_{2A}$ and  5-HT$_{2C}$  (PDB 6A93, 6BQG for 5-HT$_{2A}$ and  5-HT$_{2C}$, respectively). $\textbf{m}$ Binding site sequence similarity between 5-HT$_{2A}$ and 5-HT$_{2C}$.  $\textbf{n}$ The bilinear correlations among BAs for mu inhibitors with kappa, and delta receptors, reveal the binding site similarity and network connectivity among three proteins.   $\textbf{o}$
3D structure alignment (PDB 5C1M, 4N6H, 4DJH for mu, delta, and kappa receptors, respectively). $\textbf{p}$ 2D binding-site sequence alignment. 
$\textbf{q}$-$\textbf{t}$ Four typical cases of cross-target BA prediction of drug repurposing potentials, where some weak inhibitors of their designated targets are predicted to have high BAs to other proteins. In each chart, the two yellow frames outline the BA domains with repurposing potential, which implies compounds that have the potency to one target (BA values of $<$-9.54 kcal/mol) and do not have strong side effects on the other target (BA value of $>$-9.54 kcal/mol). }\label{cross-prediction}
\end{figure}

 Fig.  \ref{cross-prediction}$\textbf{a}$ shows the heatmap of cross-target BA predictions for 11 proteins. The diagonal elements  ($n,n$) are the squared Pearson correlation coefficients ($P^2$s) of the 10-fold cross-validation (CV) of the ML BA predictions (ML-BAs) for the corresponding protein inhibitor data sets. The minimum $P^2$ is 0.615 for catB-ext dataset and the maximum $P^2$ is 0.794 for delta-ext dataset, which suggests that 11 models have excellent model accuracy and are very reliable for cross-target predictions. Interestingly, the catB-ext dataset has 2285 compounds, while the delta-ext dataset has 6338 compounds, indicating that larger datasets lead to more reliable machine learning models than small datasets do. 
The details of $P^2$ for 11 targets could be found in Supplementary Table 9 with the consensus method using BT-FPs fingerprints.

The off-diagonal elements in Fig.  \ref{cross-prediction}$\textbf{a}$ show cross-target  predictions. The element ($n,m$), $n\neq m$,  in the heatmap  is the highest BA value  of the $n$th dataset  predicted by the $m$th  model. For example, element (7,9) is the highest BA value out of 2285 values of the catB-ext dataset predicted by the delta-ext dataset ML model established by 6338 compounds. 
The higher the highest predicted BA value is  (i.e., the darker the color is), the stronger the side effect or higher repurposing potential is. Therefore,  some 5-HT$_{2A}$,  5-HT$_{2C}$, and D$_2$ inhibitors may bind strongly to 5-HT$_{6}$.  Additionally, 
some delta and kappa inhibitors are predicted to have strong effects on the mu receptor. These effects can be either off-target side effects or repurposing potential, depending on the inhibitors' BAs for their original targets.    On the other hand, Fig.  \ref{cross-prediction}$\textbf{a}$ also reveals how safe a set of drug candidates is for other targets. For example, catB inhibitors studied in this work may not pose any side-effect concern to 5-HT$_{2A}$, 5-HT$_{2C}$, 5-HT$_{6}$, D$_2$, NMDA, NOP, catL, delta, kappa, and mu proteins.  

Figures  \ref{cross-prediction}$\textbf{b}$-$\textbf{j}$ give  some typical examples of side effects detected by cross-target BA predictions. In each chart, one target and two proteins are involved. The colors on the data points show the experimental BAs (kcal/mol) for the target in the title. The $x$- and $y$-axes of each data point are the predicted BAs by using ML models of the other two proteins, respectively. The light blue background color indicates a side-effect free zone judged by the threshold value -9.54 kcal/mol. For example,   Fig.  \ref{cross-prediction}$\textbf{b}$  plots many experimentally potent 5-HT$_{2C}$  inhibitors. Based on our ML models, none of them may induce any serious side effect on NMDA   ($x$-axis) or catL proteins ($y$-axis).  It is interesting to note from Fig.  \ref{cross-prediction}$\textbf{c}$  that most D$_2$  inhibitors are not potent and they are weak binders to NMDA  and catB proteins, according to our predictions.     However, Fig.  \ref{cross-prediction}$\textbf{e}$  and Fig.  \ref{cross-prediction}$\textbf{i}$ shows that most NOP  inhibitors are very potent  and   bind strongly with delta   and mu receptors.

Our cross-target analysis can be applied to identify not only compound-target interactions but also target-target interactions. For example, Fig.  \ref{cross-prediction}$\textbf{k}$ shows that the catB  inhibitors bind nearly linearly to    5-HT$_{2C}$ and 5-HT$_{2A}$ receptors, which are crucial for modulating drug-addictive behaviors \cite{robinson2008opposing} (A similar linear relation can be found in Fig.  \ref{cross-prediction}$\textbf{i}$.). Strong 5-HT$_{2A}$ binders are also strong  5-HT$_{2C}$ binders for catB inhibitors.  The correlation between their predicted BAs is as high as $P=0.625$. 
This correlation indicates that  5-HT$_{2C}$ and 5-HT$_{2A}$ receptors are strongly correlated, which is confirmed by the 3D protein structure alignment (Fig.  \ref{cross-prediction}$\textbf{l}$) and 2D sequence alignment of their binding sites (Fig.  \ref{cross-prediction}$\textbf{m}$), respectively. The structure alignment shows  5-HT$_{2C}$ and 5-HT$_{2A}$ receptors have the same global structure. The binding site sequence identity of these two proteins is as high as 78.7\%. 

 A more interesting relationship is the drug-mediated bilinear correlation among the three targets.  Fig.  \ref{cross-prediction}$\textbf{n}$ displays an interesting example.  The ML-predicted BAs of mu-ext inhibitors not only linearly correlate to delta and kappa receptors but also have their experimental BA values (shown in colors) linearly correlated with those of delta and kappa receptors. Therefore, a strong mu inhibitor is simultaneously a strong delta and kappa binder.   This bilinear relationship is confirmed by the 3D alignment of three proteins, i.e., mu, delta, and kappa receptors in  Fig.  \ref{cross-prediction}$\textbf{o}$. The 2D sequence alignment among mu, delta, and kappa receptors in  Fig.  \ref{cross-prediction}$\textbf{p}$  show minor differences in the three receptors. These high similarities reveal the structural basis for the unveiled drug-mediated bilinear target relationship from our ML models.    



Figures \ref{cross-prediction}$\textbf{q}$-$\textbf{t}$ depict repurposing potentials,   which are evaluated by our cross-target BA predictions. In each chart of Fig.  \ref{cross-prediction}$\textbf{q}$-$\textbf{t}$, inhibitors that are inactive to their own target may be predicted to be effective binders of other proteins. For example, Fig.  \ref{cross-prediction}$\textbf{r}$, some compounds that are experimentally inactive for the delta receptor have very high ML-BAs (i.e., BA values $<$-9.54 kcal/mol) both on D$_2$ and kappa receptors, which means that these inactive delta inhibitors are possible inhibitors for D$_2$ and kappa receptors. Similarly, in Fig. \ref{cross-prediction}$\textbf{s}$ and Fig. \ref{cross-prediction}$\textbf{t}$, many inactive DAT\_binding and all-DAT\_uptake inhibitors are predicted to be potent to D$_2$ and kappa receptors. Additional analysis of side effects and repurposing can be found in Supplementary Fig. 2. 

Figures \ref{cross-prediction}$\textbf{i}$ and \ref{cross-prediction}$\textbf{n}$
indicate that it is possible to develop drugs to simultaneously inhibit four major opioid receptors, delta, kappa, mu, and NOP (Nociceptin). A trilinear relationship is already unveiled in  Fig. \ref{cross-prediction}$\textbf{n}$, and NOP share many potent inhibitors with delta and kappa receptors as shown in Fig. \ref{cross-prediction}$\textbf{i}$. Supplementary Fig. 2 suggests that delta, kappa, and mu inhibitors may have serious side effects with D$_2$ and 5-HT$_{2A}$. 
\subsection{Evaluation of existing anti-addiction medications}

The mechanism of opioid dependence is very complicated and mainly about the reward loop of the brain and neurotransmitters, like dopamine and endorphin. Table \ref{table_medication} gives the binding affinity values (kcal/mol) predicted by four key opioid targets (namely, NOP, mu, kappa, and delta), two DAT targets (all\_DAT\_binding and all\_DAT\_uptake), and two hERG targets (hERG\_binding and hERG\_clamp) by different ML models for 12 existing anti-addiction medications. The first nine medications (i.e., Buprenorphine, Methadone, Naltrexone, Naloxone, Dihydrocodeine, Tapentadol Hydrochloride, Oxymorphone Hydrochloride, Morphine Sulfate, and Fentanyl) target opioid receptors, and the last three ones (Rimcazole, Modafinil, and Ibogaine) were designed for cocaine addiction. The number in the bracket is the maximal similarity score in the range of 0 and 1 determined via the Tanimoto coefficient between a medication and molecules collected for each target. From the viewpoint of machine learning, a higher similarity means a more reliable prediction.   

\begin{table*}[h]\Large
	\centering
     \caption{Binding affinity values (kcal/mol) predicted by eight ML models for 12 existing anti-addiction medications. The number in the bracket is the similarity score. }\label{table_medication}
    \resizebox{0.9\hsize}{!}{
    \begin{tabular}{ccccccccc}
        \hline
        Medication & NOP-ext & mu-ext & kappa-ext & delta-ext & all\_DAT\_binding & all\_DAT\_uptake & hERG\_binding & hERG\_clamp \\
        \hline
        Buprenorphine &-9.4 (1.00)  &	-12.5 (1.00)  & -12.9 (1.00) &-11.6 (1.00)&	-9.0 (0.83)&	-8.7 (0.83)&	-7.4 (0.85)&	-6.9 (0.88)\\
        Methadone&	-9.2 (0.69)&	-11.8 (1.00)&	-9.0 (1.00)&	-8.5 (1.00)&	-8.6 (0.80)&	-8.7 (0.80)&	-7.7 (0.79)&	-7.0 (0.90)\\
        Naltrexone & -9.9 (0.77)&	-12.6 (1.00)&	-12.1 (1.00)&	-10.5 (1.00)&	-8.7 (0.82)&	-8.3 (0.83)&	-7.4 (0.81)&	-6.9 (0.86)\\
        Naloxone & -10.3 (0.70)&	-11.5 (1.00)&	-11.0 (1.00)&	-9.8 (1.00)&	-8.5 (0.83)&	-8.4 (0.82)&	-7.3 (0.81)&	-7.0 (0.88)\\
        Dihydrocodeine&	-11.1 (0.76)&	-9.4 (0.98)&	-7.9 (0.98)&	-7.4 (0.98)&	-8.8 (0.84)&	-8.4 (0.84)&-7.6 (0.84)&-7.1 (0.89)\\
         \makecell{Tapentadol\\ Hydrochloride } & -10.0 (0.63)&-9.5 (0.93)&-9.1 (0.74) &-8.5 (0.70)&-8.8 (0.80)&-9.0 (0.79) &-8.7 (0.78) &-7.8 (0.82)\\
        \makecell{Oxymorphone\\ Hydrochloride} &-10.5 (0.70) & -11.5 (0.98) & -10.0 (0.98) & -9.7 (0.98) &-8.8 (0.83) &-8.9 (0.82) &-8.5 (0.83) &-7.6 (0.84)\\
        \makecell{Morphine\\ Sulfate} & -10.7 (0.74) & -11.2 (0.97) & -9.8 (0.97) & -9.5 (0.97) & -8.4 (0.82) & -8.8 (0.79) & -8.9 (0.81) & -6.8 (0.81) \\
        Fentanyl& -9.2 (0.84) & -12.5 (1.00) & -9.1 (1.00) & -9.6 (1.00) & -8.8 (0.81) & -8.4 (0.81) & -8.0 (0.79) & -7.8 (0.89)  \\
        Rimcazole&	-10.0 (0.67)&	-9.6 (0.69)&	-8.7 (0.69)&	-8.6 (0.69)&	-8.8 (0.81)&	-8.8 (0.80)&	-8.1 (0.81)&-7.9 (0.90)\\
        Modafinil&	-9.1 (0.55)&	-8.3 (0.56)&	-7.7 (0.57)&	-7.9 (0.56)&	-8.0 (0.78)&	-8.0 (0.77)&	-7.7 (0.78)&	-6.9 (0.87)\\
        Ibogaine&-10.8 (0.70)&	-9.5 (0.79)&	-8.9 (0.79)&	-8.8 (0.79)&	-7.9 (0.85)&	-7.6 (0.83)&	-8.1 (0.83)&	-7.8 (0.89)\\
        \hline
    \end{tabular}
    }
\end{table*}

Buprenorphine is one of a few FDA-approved drugs for chronic pain, acute pain, and opioid use disorder.  It has a high experimental binding affinity to the mu receptor (-12.0 kcal/mol), but relatively low binding affinities with other opioid receptors. Therefore,  it is described as a partial agonist and believed to only partially activate opiate receptors. Similar to the trilinear relation revealed in Fig. \ref{cross-prediction}$\textbf{n}$,  Buprenorphine is also an antagonist for kappa and delta receptors with binding affinities of -11.2 kcal/mol and -11.7 kcal/mol, respectively \cite{khanna2015buprenorphine}. These results are highly consistent with the predicted BAs of -12.5, -12.9, and -11.6 kcal/mol by our ML models in Table \ref{table_medication} for mu, kappa, and delta, respectively. From the table, one can also find that buprenorphine strongly binds to the NOP receptor with predicted BAs -9.4 kcal/mol and weakly binds to DAT receptors with predicted BAs -9.0 and -8.7 kcal/mol, respectively. Particularly, since hERG is a primary side effect concern for novel medications, the side effect threshold to hERG was set to -8.18 kcal/mol ($K_i$ =1 $\mu$M). We find that buprenorphine hardly has any hERG side effect, since the predicted hERG BAs are as low as -7.4 and -6.9 kcal/mol for hERG\_binding and hERG\_clamp, respectively. Additionally, as buprenorphine blocks these opioid receptors, it can decrease the effect of any subsequent opioid use. It also exhibits slower dissociation from the mu receptor compared with other opioids, which may contribute to prolonged analgesia and less potential for withdrawal when used appropriately for chronic pain. Other similar agonist drugs, such as methadone and levomethadylacetate, also known as levo-$\alpha$-acetylmethadol (LAAM) can be used for maintenance treatment against opioid dependence \cite{veilleux2010review}. 

Methadone is another FDA approved medication for opioid dependence. It is a full agonist, meaning that it can occupy  key opioid receptors, such as  NOP, mu,  kappa, and delta receptors, which is verified in our ML models as shown in Table \ref{table_medication}. Indeed, its predicted BAs for these  four receptors are  -9.2, -11.8, -9.0, and -8.5 kcal/mol, respectively. As such,  methadone reduces the painful symptoms of opiate withdrawal and blocks the euphoric effects of other opioid drugs.  Compared to heroin and other opioid agonists used for non-medical purposes, methadone has a lasting effect and  prevents the occurrence of the frequent peaks and valleys associated with compulsive behaviors. Moreover, methadone only weakly bind to DAT and has little effect on hERG (the predicted BA values for four models are -8.6, -8.7, -7.7, and -7.0 kcal/mol, respectively). Our results in Table \ref{table_medication} indicate that methadone is less effective than buprenorphine for treating opioid dependence.  
 
Opioid antagonists are another class of medications for opioid dependence.   An example is a naloxone or naltrexone, an FDA-approved drug for counteracting opioid overdoses. These antagonists bind to opioid receptors but exert no direct influence, either excitatory or inhibitory, on the post-synaptic cell. They successfully block opioid receptors, rendering subsequent opioid ingestion ineffective, and thus precipitating opioid withdrawal for dependent individuals \cite{veilleux2010review}. As shown in Table \ref{table_medication}, both naloxone and naltrexone have high mean predicted BAs (around -10.7 and -11.3 kcal/mol) or strong effects on four receptors (i.e., NOP, mu, kappa, and delta), respectively, and they have weak side effects on DAT (predicted BA values are over -9.54 kcal/mol), and has no side effects on hERG (predicted BA values are larger than -8.18 kcal/mol), which is similar with that of buprenorphine and methadone. The predicted BAs on different important opioid receptors verified that these approved medications could be potent in the treatment of opioid addiction.

Dihydrocodeine is often prescribed as an alternative to methadone or buprenorphine for pain or severe dyspnea, or as an antitussive, either alone or compounded with paracetamol (acetaminophen) (as in co-dydramol) or aspirin. It is a semi-synthetic opioid analgesic and exists in both extended-release and immediate-release form, which is confirmed by the predicted BA values for four key receptors (i.e., -11.1, -9.4, -7.9, and -7.4 kcal/mol for NOP, mu, kappa, and delta, respectively), DAT and hERG in Table \ref{table_medication}. As an opioid agonist, dihydrocodeine could be used as the second line of treatment. A 2020 systematic review found low-quality evidence that it may be no more effective than other routinely used medications in reducing illicit opiate use \cite{carney2020dihydrocodeine}. Our results in Table \ref{table_medication}  support this view. 

Extended-release, long-acting (ER/LA), and immediate-release (IR) opioid analgesics, such as Tapentadol Hydrochloride, Xylophone Hydrochloride, Morphine Sulfate, and Fentanyl in Table \ref{table_medication}, are powerful pain-reducing medications that have both benefits as well as potentially serious risks. For instance, the predicted BA values on hERG\_binding are as high as -8.7, -8.5, -8.9, and -8.0 kcal/mol for these four medications, respectively, which suggests serious heart failure concerns. Moreover, these four medications have moderately high predicted binding affinities on mu, kappa, delta, and NOP receptors, indicating their capacity for pain mitigation. For example, Fentanyl is a synthetic opioid that is 50-100 times stronger than morphine (from DEA, United States Drug Enforcement Administration). Pharmaceutical fentanyl was developed for the pain management treatment of cancer patients, and applied in a patch on the skin. The experimental BA values for mu, kappa, and delta are around -12.8, -8.2, and -9.0 kcal/mol \cite{raynor1994pharmacological}, which are consistent with our predicted BA values -12.5, -9.1, and -9.6 kcal/mol, respectively. 

The mechanism and potential treatment of other drug addictions, like cocaine addiction is much more complicated than those of opioid addiction. Currently, no cocaine addiction medication has been approved by FDA. The details of analysis and discussion about cocaine addiction medication can be found in the literature \cite{Gao2021}. Table \ref{table_medication} lists three potential medications for cocaine addiction, namely rimcazole, modafinil, and ibogaine, which are not effective for four opioid targets, namely, NOP, mu, kappa, and delta, as they could only partially block these targets. For instance, rimcazole and ibogaine can strongly bind to NOP and mu receptors with predicted BA values smaller than or close to -9.54 kcal/mol, but are not effective on kappa and delta receptors.

Rimcazole, as a DAT reuptake inhibitor, can reduce the effects of cocaine by binding to DAT, and was reported to have a BA value of -9.54 kcal/mol to DAT \cite{newman2021new}. However, we obtain a lower predicted BA (i.e., -8.8 kcal/mol) for rimcazole, which suggests that its effect on DAT is questionable. Additionally, the potential heart-related side effect is supported by the predicted BA value (-8.1 kcal/mol) for hERG\_binding in Table \ref{table_medication}, which is close to the threshold value -8.18 kcal/mol. Rimcazole is not pursued anticocaine addiction now.  

Modafinil, as an atypical DAT inhibitor, is a potential treatment for cocaine addiction and has been approved for the treatment of excessive sleepiness. Especially, it has a low DAT affinity (-6.9 kcal/mol) lower than that of cocaine and still prevents DAT from being blocked by cocaine \cite{newman2021new}. In Table \ref{table_medication}, the predicted BA values for modafinil to all\_DAT\_binding and all\_DAT\_uptake are about -8.0 kcal/mol. Additionally, modafinil has a low predicted BA to hERG, suggesting no serious hERG risk. Note that the reliability of our ML predictions for Modafinil is relatively low due to its low similarity to existing molecules.

Ibogaine is also an atypical inhibitor of DAT with a BA value of -7.8 kcal/mol \cite{efange1998modified}. Importantly, its severe side effects and related death are of serious concern \cite{koenig2015anti}. As shown in Table \ref{table_medication}, our ML model predicted high BA values of ibogaine to hERG (as high as -8.1 kcal/mol on hERG\_binding), which confirms that ibogaine could have a high risk for heart failure.

The above analysis of 12 existing anti-addiction medications and the predicted binding affinity values in Table \ref{table_medication} sheds light on the medication discovery of opioid or cocaine addiction. Additionally, the overall high similarity scores given in the table ensure the reliability of our ML predictions. 

\subsection{ Impact of persistent Laplacian-based fingerprints }

Although many self-supervised learning-based molecular fingerprints, including our BT-FPs/BT$_f$-FPs,  have demonstrated a better performance than conventional fingerprints, which are often short of geometric information about molecules.  
For example, some molecules share the same canonical SMILES  but may have different chemicals and/or biological properties, like Trans-1,2-Dichlorocyclohexane and Cis-1,2-Dichlorocyclohexane (see Fig.   \ref{bells}$\textbf{e}$). To address this challenge,  we introduce the persistent Laplacian in combination with our TIDAL architecture to retain the stereochemical and physical information. Additionally, we set the total dimension of molecular fingerprints TIDAL-FPs/TIDAL$_f$-FPs to 512 after feature fusion, and thus we do not need to adjust the bidirectional transformer framework to achieve the optimal learning results for different problems.

Supplementary Fig.   3$\textbf{a}$ plots the prediction results of different combinations of fingerprints integrating with EANN on 11 drug addiction-related classification datasets using the TIDAL framework. It is seen that the persistent Laplacian-based fingerprints PL-FPs outperform   BT-FPs and BT$_f$-FPs in 7/11 and 9/11 of the cases, respectively. The AUC values of combined molecular fingerprints TIDAL-FPs are  1.7$\%$ and 6.1$\%$ higher than those of BT-FPs and BT$_f$-FPs on average over all cases, respectively.  Especially, 
TIDAL-FPs outperform BT$_f$-FPs by 16.4$\%$   in AUC on the 5-HT$_{2A}$ dataset. The details of AUC values on drug addiction-related classification datasets are listed in Supplementary Table 6. Therefore, these results validate that the stereochemical information exacted from the persistent Laplacian method indeed helps the downstream prediction tasks.  The fusion of PL-FPs and BT-FPs/BT$_f$-FPs improves the accuracy and stability of ML prediction. Furthermore, mathematics-based molecular fingerprints complement to sequence-based fingerprints.   

Additionally, we analyze prediction errors for molecules with trans-cis, chiral or steric effects in Supplementary Table 7 to reveal the need for PL-FPs. For instance, for the classification task, there are two molecules with the same molecule formula $\rm{C_{24}H_{36}N_4O_2}$ and SMILES in the  Mu dataset. However, they have different geometries,  i.e., trans and cis isomers. The probabilities of classification with PL-FPs and BT-FPs are 0.727 and 0.676 for the trans molecule, respectively. The errors in probability are 29.7\% and 32.4\%, respectively, which means that although the predicted label is correct for both fingerprints,  PL-FPs have a higher probability and smaller error than BT-FPs.   Similarly, for the regression task, PL-FPs achieve smaller errors than BT-FPs for the molecules with the same SMILES in the hDAT\_binding dataset. In the last second and fourth columns of the table, the predicted value and true value are listed, and the error is in the percent of the difference between them over the true value. 

Supplementary Table 8 gives the comparison results between the persistent Laplacian (PL) method and traditional persistent homology (PH) on various datasets. Here, we consider two commonly used complexes, the Rips complex and Alpha complex in the calculation of PH. For classification tasks, the PL method achieves higher AUC values than PH on 9/10 datasets except for the 5-HT$_{6}$ dataset. For regression tasks, PL gets a higher squared Pearson correlation coefficient ($P^2$) and smaller RMSE than PH on all 4 datasets. This is other evidence that PL proposed in the present work outperforms PH on molecular property predictions.

\subsection{Predictive power of bidirectional transformer-based fingerprints} 

In this work, we take the self-supervised learning (SSL) strategy in the fine-tuning stage and the workflow of the SSL strategy can be found in Supplementary Fig.  4. During the fine-tuning phase, we use the task-specific dataset with labels as input data. The labeled data promote the model to extract task-related information from the training data, which may improve the accuracy of downstream tasks. The molecular fingerprints generated from the fine-tuned model are called BT$_f$-FPs. On the other hand, the pre-trained model itself is generated based on the syntax information of SMILES and thus can also be utilized directly to generate molecular fingerprints named BT-FPs.

\begin{figure*}[!tpb]
\centering
\includegraphics[width=15cm]{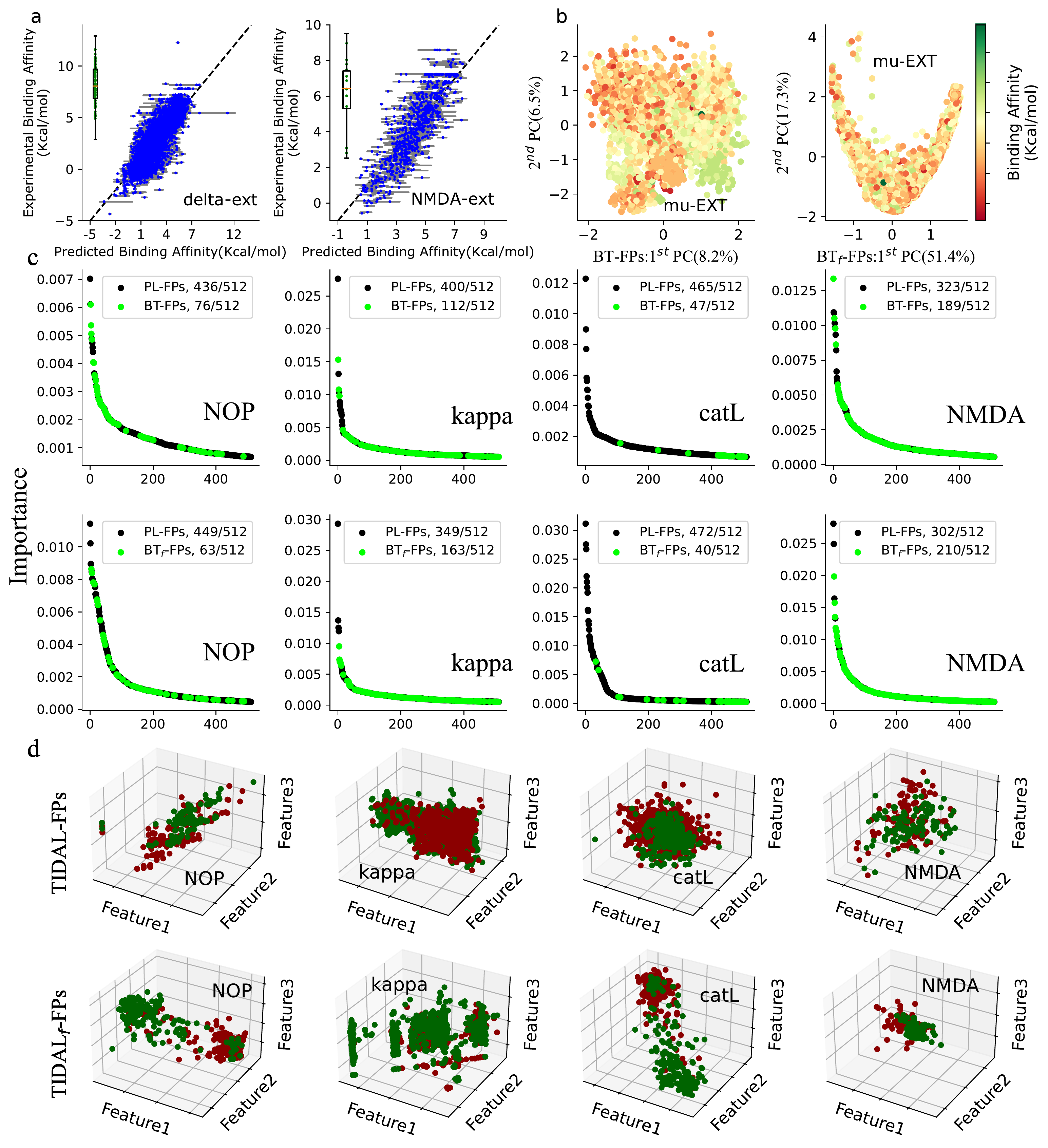} 
\caption{Results from the TIDAL framework and feature analysis on molecular activity and binding affinity of drug addiction-related datasets. $\textbf{a}$ Prediction results of BT-FPs with GBDT algorithm on delta-ext (left) and NMDA-ext (right) datasets. The box plots statistic $P^2$ values for 633 (left) and 81 (right) independent samples examined over 10 independent machine learning experiments. $\textbf{b}$ The variance ratios in the first two components from the principal component analysis (PCA) for BT-FPs (left) and BT$_f$-FPs (right) on the mu-ext dataset. $\textbf{c}$ The feature importance ranking of TIDAL-FPs (top panel)/TIDAL$_f$-FPs (bottom panel) on molecular binding affinity datasets, including NOP, kappa, catL, and NMDA. For four datasets, the most important features of TIDAL-FPs/TIDAL$_f$-FPs are from PL-FPs except the NMDA dataset with TIDAL-FPs. $\textbf{d}$ Visualization of top three important features of TIDAL-FPs (top panel) and TIDAL$_f$-FPs (bottom panel) on the NOP, kappa, catL, and NMDA datasets.}\label{main_results}
\end{figure*}

Figure   \ref{main_results}$\textbf{a}$ shows the prediction performance with BT-FPs on molecular binding affinity datasets delta-ext and NMDA-ext. The gray bar at each point is the deviation of predicted binding affinity with 10 repetitions with different random seeds. We find that the delta-ext dataset has good prediction results with $P^2=0.780$. For the NMDA-ext dataset, there are only 652 molecules used to train the model consisting of BT-FPs and EANN algorithm, leading to the fluctuation in the prediction with $P^2=0.725$. Similar situations are found in other molecular binding affinity datasets as shown in Supplementary Fig.  5.  

In Supplementary Fig.  3$\textbf{a}$, we find that bidirectional transformers framework with SSL strategy on task-specific task performs better than that without SSL strategy, that is, the TIDAL$_f$-FPs achieve higher AUC values than TIDAL-FPs for all 11 datasets. In particular, the SSL strategy performs better on small datasets than that on big datasets. For instance, NMDA and NOP datasets have only  246 and 431 molecules in which 80$\%$ molecules are used for the training set, and AUC values with TIDAL$_f$-FPs compared to those with TIDAL-FPs are increased up to 1.9$\%$ and 2.7$\%$, respectively, which are both larger than 1.3$\%$ the average improvement for the left large tasks.  Conventional methods usually cannot extract enough information from such a small dataset to obtain satisfactory results. In our TIDAL model, the pre-training module with deep bidirectional transformers enables the model to capture general information about molecules, and the fine-tuning process supplies additional task related knowledge about molecules. We further complement this information with topological and geometric information generated from persistent Laplacian to ultimately boost the generalization ability of the TIDAL model and raise the performance on the NMDA and NOP cases with AUC values 0.987 and 0.985, respectively.   

In Supplementary Fig.  3$\textbf{b}$ and Fig.  3$\textbf{c}$, we compare the performance of binding affinity predictions on 11 regression tasks of opioid datasets with and without fine-tuned process corresponding to BT$_f$-FPs (see Supplementary Fig.  3$\textbf{b}$) and BT-FPs (see Supplementary Fig.  3$\textbf{c}$), respectively, and find that BT-FPs integrating with GBDT, DNN, and consensus method have better performance than BT$_f$-FPs for all datasets with 16.4$\%$, 12.2$\%$, and 11.0$\%$ average increasing of $P^2$, respectively. Especially, with the GBDT algorithm, the value of $P^2$ for the NOP-ext dataset is increased up to 34.5$\%$ from 0.478 to 0.643 obtained by BT$_f$-FPs and BT-FPs, respectively. This result is also consistent with the performance on classification tasks in Supplementary Fig.  3$\textbf{a}$ where BT-FPs result in higher AUC values than BT$_f$-FPs in 9/11 cases. Additionally, regardless of BT-FPs and BT$_f$-FPs, DNN has more advantages over GBDT on big datasets. For BT-FPs, the values of $P^2$ with DNN are larger than those with GBDT in 9/11 tasks, except for two small datasets, namely NMDA-ext with 815 samples and NOP-ext with 2063 samples. For BT$_f$-FPs, DNN has better prediction performance than GBDT in all tasks. As the consensus method takes the average predictive results of GBDT and DNN, it obtains the best performance in most cases (9/11) with BT-FPs and in all cases with BT$_f$-FPs. The complete results with $P^2$, RMSE, and MAE on regression tasks are listed in Supplementary Table 9.

The proposed TIDAL-FPs are fingerprints containing stereochemical information of molecules, which are a projection of different physical and chemical information of molecules. A more detailed analysis of TIDAL-FPs can be found in Supplementary Note 3.

\begin{figure}[!tpb]
\centering
\includegraphics[width=15cm]{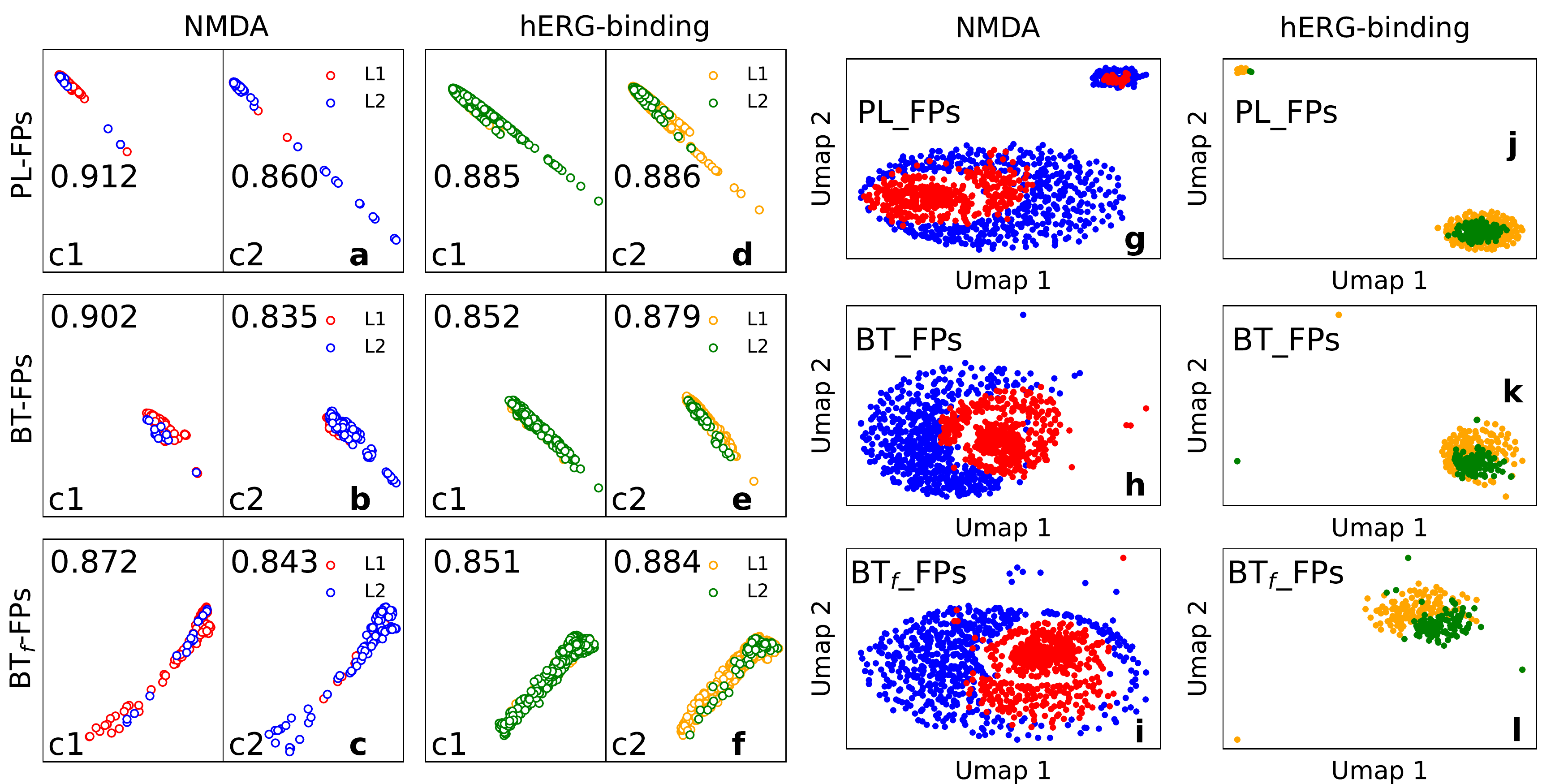}
\caption{Visualization of NMDA and hERG\_binding datasets with various fingerprints, like PL-FPs, BT-FPs, and BT$_f$-FPs by RS plot and Umap, respectively. $\textbf{a}$-$\textbf{f}$: each section means a class one (c1) or two (c2) with labels L1  or L2, respectively, and the samples are colored according to their predicted labels from GBDT algorithm. The $x$ and $y$ axes indicate the residue and similarity scores, respectively. $\textbf{g}$-$\textbf{l}$: clustering visualizations by Umap for PL-FPs, BT-FPs, and BT$_f$-FPs on NMDA and hERG\_binding datasets. }\label{rs_plot}
\end{figure}

\subsection{Residue-Similarity (R-S) scores }

 In this section, we carry out  Residue-Similarity (R-S) analysis  \cite{hozumi2022ccp}  for the clustering visualization of performance with different fingerprints, like PL-FPs, BT-FPs, and BT$_f$-FPs. For classification problems with 2 classes, the traditional methods, like receiving operation characteristic (ROC) curve and Area Under the ROC Curve (AUC) curve can do similar work as R-S scores, however, R-S scores can be applied to an arbitrary number of classes, which is its advantage over the traditional methods.

Figures  \ref{rs_plot} $\textbf{a}$-$\textbf{f}$ show the R-S plot of the NMDA and hERG\_binding datasets with different fingerprints, including PL-FPs, BT-FPs, and BT$_f$-FPs, where the R-S scores are represented as the $x$ and $y$ axes, respectively, and the accuracy values of classification for each class are marked inside the plot. The left and right sections in each panel correspond to the 2 classes, class one (c1) with the label (L1) (active), and class two (c2) with the label (L2) (inactive). For the NMDA dataset, in panel (a) with PL-FPs, we find that samples both with L1 and L2 having low residue scores are more likely to be mislabeled, and the classification accuracy for L1 is higher than that for L2. In panels (b) and (c) with BT-FPs and BT$_f$-FPs, respectively, we find that samples with low similarity scores tend to be mislabeled, and the classification accuracy with PL-FPs is higher than that with BT-FPs or BT$_f$-FPs. These findings suggest that the samples with low R-S scores prefer to be misclassified, and samples with PL-FPs have higher classification accuracy than those with BT-FPs and BT$_f$-FPs, which is also consistent with the results of Supplementary Fig.  3$\textbf{a}$. In panels (d)-(f), similar results are found on hERG\_binding dataset.

Figures   \ref{rs_plot} $\textbf{g}$-$\textbf{l}$ give the visualizations by UMAP method \cite{https://doi.org/10.48550/arxiv.1802.03426} on NMDA and hERG\_binding datasets. For the NMDA dataset, there is a good clustering visualization for two classes marked by blue and red colors except for a few points distributed around. For PL-FPs, the prediction accuracy values for the classification of class c1 and c2 are 0.736 and 0.950, respectively, which are comparable with those of 0.912 and 0.860 by RS plot, however, the whole accuracy value is 0.843 for the two classes by UMAP is smaller than that by RS plot 0.886. For BT-FPs and BT\_-FPs, the accuracy values for c1 and c2 are 0.744 and 0.926,  and 0.832 and 0.860, respectively. The whole accuracy values are 0.835 and 0.846, which are both smaller than 0.869 and 0.858 from the RS plot. These findings suggest that though UMAP has a good clustering visualization, RS plot has a better prediction performance than UMAP. Similar results are found in the case of hERG\_binding datasets.

\section{Methods}

\subsection{Drug addiction related, DAT, and hERG datasets}

The proposed TIDAL model and its results for molecular activity and binding affinity prediction involve  22 drug addiction-related datasets, 4 hERG datasets, and 12 DAT datasets. Detailed information on these datasets and the CheMBL dataset used in the pre-training are given in Supplementary Table 10 and Table 11. More descriptions of  the datasets can be found in Supplementary Note 1.

We implement our final predictions by EANN, and compare the results with other standard ML algorithms, like logistic regression (LR), linear support vector machines (LSVM), support vector machines with a radial basis function (RBF) kernel (SVM), and neural network (NN). In order to eliminate systematic errors in the ML predictions, we take average values of 10 realizations of the results of the models. Additionally, the method of consensus used on datasets refers to the average predicted values from different ML models for each molecule. In this work, we use the area under the receiver operating characteristic convex hull (AUC-ROC), Matthews correlation coefficient (MCC), accuracy, sensitivity, specificity, and F$_1$ score to evaluate the performance of the classification model, while we use the squared Pearson correlation coefficient ($P^2$), root-mean-square error (RMSE), the coefficient of determination ($R^2$), and mean absolute error (MAE) to assess the performance of regression task. All definitions of these metrics can be found in Supplementary Note 2 and the hyperparameters of the ML algorithm are provided in Supplementary Table 12 and Table 13. Moreover, we applied the repeated tenfold cross-validation method to assess the generalization of prediction models.
  
For 22 drug addiction related datasets, they are collected from literature \cite{warszycki2021pharmacoprint} and ChEMBL database \cite{gaulton2012chembl}, including 11 classification tasks and 11 regression tasks. As ChEMBL contains numerical values of particular parameters that determine the activity of the compounds, molecules are only considered in our model if their activities and binding affinity are quantified by $K_i$, $pK_i$ or $\rm{IC}_{50}$ and have been tested in human protein assays. The $\rm{IC}_{50}$ values are approximately converted to $K_i$ by expression $K_i=\rm{IC}_{50}/2$. For each dataset, we consider all the compounds with $K_i$ or $\rm{IC}_{50}$ values but delete redundant ones and the label we used for training and testing is the binding affinity (1.3633*$\rm{log_{10}}$$K_i$) in regression task. For classification task, fetched compounds are classified to actives ($pK_i$ or equivalent $>$ 7) and inactive ($pK_i$ or equivalent $<$ 6) \cite{smieja2016average}. Note that we distinguish the regression and classification task by the term ``ext", like 5-HT$_{2A}$ dataset is for classification, and 5-HT$_{2A}$-ext is for regression. Term ``ext" also means no limitation about the value of $K_i$, $pK_i$ or $\rm{IC}_{50}$. The details of the datasets can be found in Supplementary Table 10.

Additionally, 4 hERG and 12  DAT datasets are adopted from the literature \cite{lee2021toward}, where the filters were developed to sort the data based on how they were acquired, that is, from either patch-clamp electrophysiology (referred to as ``clamp") or radioligand binding assays ``binding"), or inhibition of dopamine uptake (``uptake"). As expected, in the DAT dataset, the majority of the data are of human DAT (hDAT) and rat DAT (rDAT) since most experiments were carried out with hDAT heterogeneously expressed in vitro cell lines or with rat brain tissues. So the all-DAT dataset includes hDAT, rDAT, and a few other species. 

\subsection{Persistent Laplacian-based molecular fingerprints (PL-FPs)}

Graph theory, including geometric graph theory, algebraic graph theory, and topological graph theory, focuses on the relationship among nodes, edges, faces, and their high-dimensional extensions. 
Mathematically, combinatorial graph theory (CGT) is intrinsically related to simplicial complexes and algebraic topology. One of the important development of CGT is the formulation of topological  Laplacians (TLs), which offer topological information in its harmonic spectra and geometric information, including algebraic connectivity, in its non-harmonic spectra. One specific TLs 
is defined on a compact Riemannian manifold based on the de Rham-Hodge theory \cite{zhao2020rham}. The harmonic part of the Hodge Laplacian spectrum gives rise to topological invariants, however, the non-harmonic part of the Hodge Laplacian spectrum provides geometric shape analysis \cite{zhao2020rham}. Evolutional de Rham-Hodge theory was developed for a family of evolving manifolds   \cite{chen2021evolutionary}. 


Recently, persistent Laplacian (PL) or persistent spectral graph (PSG) theory has been introduced to connect persistent homology and CGT \cite{wang2020persistent}. Like persistent homology, a filtration process is used to create a series of simplicial complexes, on which persistent spectral graphs are defined. 
 The persistent spectral analysis (PSA) resulting from PSGs extends topological data analysis (TDA) to geometric shape analysis. Specifically, the change in the null space dimensions of the PLs during the filtration represents the persistence of topological invariants, while the non-zero eigenvalues and associated eigenfunctions of the PLS reveal the geometric shape evolution of the data during the filtration \cite{wang2020persistent}. 
   
PL  generates a sequence of simplicial complexes by   filtration.   An oriented simplicial complex $K$ is a sequence of sub-complexes $(K_t)^m_{t=0}$ of $K$:
\begin{equation}
\emptyset = K_0 \subseteq K_1 \subseteq K_2 \subseteq \cdots \subseteq K_m = K.
\end{equation}
  Let $C_q(K_t)$ be the corresponding chain group of each subcomplex $K_t$ and 
	   $\partial_q^t: C_q(K_t) \to C_{q-1}(K_t)$ be the $q$-boundary operator.  If $0<q \le \dim K_t$, we have 
\begin{equation}
    \partial_q^t(\sigma_q) = \sum_{i}^q(-1)^i \sigma^i_{q-1}, \quad \forall \sigma_q \in K_t,
\end{equation}
where $\sigma_q = [v_0, \cdots, v_q]$ is any $q$-simplex and $\sigma^{i}_{q-1} = [v_0, \cdots, \hat{v_i} ,\cdots,v_q]$ being the oriented $(q\!-\!1)$-simplex constructed by removing $v_i$. If $q<0$, then $C_q(K_t)=\{0\}$ and $\partial_q^t$ is   a zero map \cite{kamber1987rham}.
It is important to define an adjoint operator of $\partial_q^t$ as the coboundary operator $\partial_q^{t^{\ast}}: C^{q-1}(K_t) \to C^q(K_t)$,   mapping from $C_{q-1}(K_t)$ to $C_q(K_t)$ by  the isomorphism   between cochain groups and chain groups $C^q(K_t)\cong C_q(K_t)$.
 
For simplicity, denote $C_q^t$ the chain group $C_q(K_t)$.  Using the natural inclusion map from $C_{q-1}^t$ to $C_{q-1}^{t+p}$, one defines the subset of $C_q^{t+p}$ as $\mathbb{C}_q^{t,p}$ with the boundary being in $C_{q-1}^t$ as
\begin{equation}
    \mathbb{C}_q^{t,p} \coloneqq \{ \beta \in C_q^{t+p} \mid \ \partial_q^{t+p}(\beta) \in C_{q-1}^{t}\}.
\end{equation}
On this subset, denote the $p$-persistent $q$-boundary operator by $\eth_q^{t,p}: \mathbb{C}_q^{t,p} \to  C_{q-1}^{t}$and the corresponding adjoint operator by $(\eth_q^{t,p})^{\ast}: C_{q-1}^{t}  \to  \mathbb{C}_q^{t,p}$ through the identification of cochains with chains.  We define the $q$-order $p$-persistent Laplacian operator $\Delta_q^{t,p}: C_q^t \to C_q^t$ in the filtration \cite{wang2020persistent}
\begin{equation}
    \Delta_q^{t,p} = \eth_{q+1}^{t,p} \left(\eth_{q+1}^{t,p}\right)^\ast + \partial_q^{t^\ast} \partial_q^t.
\end{equation}
The matrix representation of $\Delta_q^{t,p}$ in the simplicial basis is 
\begin{equation}
    \mathcal{L}_q^{t,p} = \mathcal{B}_{q+1}^{t,p} (\mathcal{B}_{q+1}^{t,p})^T + (\mathcal{B}_{q}^t)^T \mathcal{B}_{q}^t,
\end{equation}
where $\mathcal{B}_{q+1}^{t,p}$ is the matrix representation of $\eth_{q+1}^{t,p}$. The spectrum of $\mathcal{L}_q^{t,p}$ is given  as $\text{Spec}(\mathcal{L}_q^{t,p}) = \{\lambda_{1,q}^{t,p}, \lambda_{2,q}^{t,p}, \cdots,\lambda_{N_q^t,q}^{t,p}  \}$, where $N_q^t=\dim C_q^{t}$ is the number of $q$-simplices in $K_t$, and the eigenvalues are  ordered. Here, $\lambda_{2,q}^{t,p}$ denotes the smallest non-zero eigenvalue of $\mathcal{L}_q^{t,p}$. The Betti number is related to $q$-cycle information, and the number of zero eigenvalues of in the spectrum of $\mathcal{L}_q^{t,p}$ is the $q$th order $p$-persistent Betti number $\beta_q^{t,p}$
\begin{equation}
    \begin{aligned}
        \beta_q^{t,p} = \dim \ker \partial_q^t - \dim \im \eth_{q+1}^{t,p} = \dim \ker \mathcal{L}_q^{t,p} = \# \text{0 eigenvalues of } \mathcal{L}_q^{t,p}
    \end{aligned}
\end{equation}
In this study, we pay attention to the $0,1,2$th-order persistent Laplacians for molecular datasets. Mathematically,  $\beta_q^{t,p}$ from the null space of LPs tracks the number of independent $q$-dimensional holes in $K_t$ that are still alive in $K_{t+p}$. Therefore, it gives the same topological information as persistent homology does. However, the non-zero eigenvalues of the PLs reveal the homotopic shape evolution of data during filtration.  
 
For each molecule, PLs are constructed from element-specific combinations of the molecule. For each combination, the statistics of PL eigenvalues are used for PL-FPs.

\subsection{Deep bidirectional transformer fingerprints (BT-FPs)}

 A bidirectional transformer (BDT) utilizes a self-supervised learning process to learn the constitutional rules of chemical data. It employs an independent positional embedding scheme to deal with unlabeled molecular data, and acquires the importance of each symbol in the input sequence that is different from sequences learning models such as RNN and LSTM-based models \cite{vaswani2017attention}. Partially masked chemical symbols are used as input during the training to improve the model's ability to recover the original molecules.  Additionally, DBT uses an attention mechanism to take care of memory loss issues in sequential learning. An important trait of BDT is that it can be fully parallelized to well reduce the training time for massive data, which makes the training of a network with over 700 million unlabeled SMILES data possible \cite{chen2021extracting}.  

In the present work, the input of the deep bidirectional transformer is a molecular SMILES string, which is different from the sentences in traditional BERT for natural language processing, and the SMILES strings of molecules are not logically connected. So only the encoder of the transformer is used and only the masked learning task in the pre-learning process is kept, to mask part of the input SMILES symbols during the training process and then recover the masked symbols by training. Specifically, a SMILES string is divided into symbols, like C, H, N, O, =, etc., representing the atoms, chemical bonds, and connectivity. During the pre-training stage, a certain percentage of input symbols are selected randomly for three types of operations: mask, random changing, and no changing. The pre-training aims to learn fundamental constitutional information about molecules using the SSL method with unlabeled data. Moreover, a loss function is proposed to improve the rate of correctly predicted masked symbols in the pre-training process. Here, for each SMILES string, two special symbols, $< s >$ and $<\backslash s>$ are added and the former is used as the beginning of a SMILES string and the latter is a special terminating symbol. All symbols are embedded into input data with a fixed length. Additionally, a position embedding is also added to every symbol to denote the order of the symbol. The embedded SMILES strings are fed into the BERT architecture for further operation and the detailed process of the pre-training procedure can be found in Supplementary Fig.   10. In our study, more than 2.0 million unlabeled SMILES data from CheMBL are borrowed for pre-training, and thus not only the fundamental syntactic information of SMILES strings can be learned but also the global information of molecules is captured by the model. 

Both BT-FPs and BT$_f$-FPs are based on the pre-trained model and BT$_f$-FPs has an additional fine-tuning procedure by SSL strategy with labeled task-specific data compared to BT-FPs. The SSL-based fine-tuning is shown in Supplementary Fig.   4. Here, in our DBT, the maximal sequence length of an input SMILES string is set to 256 symbols, including the start and terminate ones. During the training, the embedding dimension of each symbol is set to 512 and the 512-dimensional vector retains the information of the whole SMILES string. In this extended 256 x 512 representation, one or multiple 512-dimensional vectors can be selected to represent the original molecule in principle. In our work, the corresponding vector of the leading symbol $<s>$ of a molecular SMILES string is chosen as the BT-FPs or BT$_f$-FPs of the molecule, and in the downstream tasks, these molecular fingerprints are used for molecular property prediction.

\subsection{Ensemble-assisted neural network (EANN)}

For the downstream machine learning module, we take two steps to get the output result. First, we combine PL-FPs and BT-FPs/$\rm{BT_f}$-FPs and rank the importance of combined fingerprints by gradient boosting decision tree (GBDT)   algorithm, and then select an optimal set of TIDAL-FPs/$\rm{TIDAL_f}$-FPs with a fixed number of components (such as 512) to feed the deep neural network (DNN) with $M$ hidden layers. Each layer has $m$ neural nodes. Second, we use ensemble learning models, i.e., gradient boosting decision tree (GBDT) and random forest algorithms,   to train a model with TIDAL-FPs/$\rm{TIDAL_f}$-FPs, respectively. After training, we add two trained weights (or probabilities) as two additional neural nodes on both sides of DNN in the last hidden layer. Hence, through the training of $m+2$ neural nodes of the last hidden layer, we obtain automatically optimized consensus on various datasets. This framework of machine learning integrates both advantages of DNN and GBDT named EANN. For traditional classification ML algorithms, before getting the predictive class for data, one needs to change the output, the probability values for different classes to class value. EANN is introduced here, however, no change is needed and the output is the class label we want, which can be seen as another advantage of the TIDAL model.

\subsection{Residue-Similarity (R-S) scores }

R-S scores were proposed in our recent work for analyzing and visualizing dimensionality reduction and classification algorithms \cite{hozumi2022ccp}.  
Assume that the data is $\{(\mathbf{x}_i, y_i) \mid \mathbf{x}_i \in \mathbb{R}^N, y_i \in \mathbb{Z}_L\}^M_{i=1}$, where $\mathbf{x}_i$ is the $i$th data entry, $y_i$ is the ground truth for the classification task, and/or the cluster label for the clustering task. Here, $N$ is the feature dimension, and $M$ is the total number of samples. Here, $L$ is the number of classes, and $y_i\in [0,1,...,L-1]$. The data $\mathcal{X}= \{\mathbf{x}_i\}_{i=1}^M$ can be divided into $L$ classes by taking $\mathcal{C}_l = \{\mathbf{x}_i \in \mathcal{X} \mid y_i = l\}$ and $\uplus_{l=0}^{L-1}\mathcal{C}_l = \mathcal{X}$.

An R-S plot has two components, i.e., residue score and similarity score. Supposing $y_i = l$, the residue score is defined as the inter-class sum of distances. For $\mathbf{x}_i$, it is defined as
\begin{equation}
    R_i:= R(\mathbf{x}_i) = \frac{1}{  R_{\max}}\sum_{\mathbf{x}_j \not\in \mathcal{C}_l} \|\mathbf{x}_i - \mathbf{x}_j\|,
\end{equation}
where $\|\cdot\|$ is the distance between a pair of vectors $\mathbf{x}_i$ and $\mathbf{x}_j$ and $\displaystyle R_{\max} = \max_{\mathbf{x}_i\in \mathcal{X}} R(\mathbf{x}_i)$ is the maximum value of the residue score. We use the Euclidean distance for $\|\cdot\|$.
The similarity score is given by taking the average intra-class score and is defined as
\begin{equation}
    S_i:=S(\mathbf{x}_i) = \frac{1}{|\mathcal{C}_l|} \sum_{\mathbf{x}_j \in \mathcal{C}_l} \left( 1-\frac{\|\mathbf{x}_i - \mathbf{x}_j\|}{d_{\max}} \right),
\end{equation}
 where $ \displaystyle d_{\max} = \max_{\mathbf{x}_i,  \mathbf{x}_j \in \mathcal{X}} \|\mathbf{x}_i - \mathbf{x}_j\|$ is the maximum value of pairwise distance of the dataset. Specially, $R(\mathbf{x}_i)$ and $S(\mathbf{x}_i)$ are between 0 and 1 after scaling for all $\mathbf{x}_i$. In general, larger $R(\mathbf{x}_i)$ means that the data is further from other classes, and larger $S(\mathbf{x}_i)$ means that the data is better clustered. The residue score and similarity score can be used to visualize each class separately. 

In R-S plot, $R(\mathbf{\mathbf{x}})$ is the $x$-axis, and $S(\mathbf{x})$ is the $y$-axis. For classification task, we define $\{(\mathbf{x}_i, y_i, \hat{y}_i) \mid x_i \in \mathbb{R}^N, y_i \in \mathbb{Z}_L, \hat{y}_i \in \mathbb{Z}_L\}_{i=1}^M$, where $\hat{y}_i$ is the predicted label for the $i$th sample, and then we repeat the above process by using the ground truth and visualize each class separately. Finally, we get the R-S score visualization of the classification by coloring the sample with the predicted label.

\section*{Data availability}
The 22 drug addiction related datasets,  4 hERG datasets, and 12 DAT dataset can be obtained from \url{https://weilab.math.msu.edu/DataLibrary/2D/} and \url{https://weilab.math.msu.edu/DataLibrary/3D/}, respectively.  

\section*{Code   availability}
The code of pre-trained model is available in the Github repository: \url{https://github.com/WeilabMSU/PretrainModels}  and the code of the calculation of persistent Laplacian-based fingerprints is available via \url{https://github.com/wangru25/HERMES}. 

\section*{Acknowledgements}
The work of Zailiang Zhu, Jian Jiang, and Bengong Zhang was supported by the National Natural Science Foundation of China under Grant No.11971367 and No.11972266. The work of Dong Chen and Guo-Wei Wei was supported in part by NIH grants  R01GM126189 and  R01AI164266, NSF Grants DMS-1721024, DMS-1761320, and IIS1900473, NASA grant 80NSSC21M0023, Bristol Myers Squibb, Pfizer, and MSU Foundation. All authors appreciate the technology assistance from Rui Wang.

\section*{Authors' contributions}
Zailiang Zhu, Yucang Cao, Bozheng Dou, Yueying Zhu, Dong Chen, and Hongsong Feng performed computational studies. Jian Jiang analyzed data, wrote the draft, and revised the manuscript. Guo-Wei Wei designed and supervised the project, revised the manuscript, and acquired funding. Jie Liu, Bengong Zhang, and Tianshou Zhou supervised the project and acquired funding.

\section*{Competing interests}
The authors declare no competing interests.
\bibliography{ref}

\begin{thebibliography}{10}

\bibitem{Cao2021}
Dan-ni Cao.
\newblock {Insights into the mechanisms underlying opioid use disorder and
  potential treatment strategies}.
\newblock {\em Br J Pharmacol}, pages 1--17, 2021.

\bibitem{Gao2021}
Kaifu Gao, Dong Chen, Alfred~J. Robison, and Guo~Wei Wei.
\newblock {Proteome-Informed Machine Learning Studies of Cocaine Addiction}.
\newblock {\em Journal of Physical Chemistry Letters}, 12(45):11122--11134,
  2021.

\bibitem{gao2020generative}
Kaifu Gao, Duc~Duy Nguyen, Meihua Tu, and Guo-Wei Wei.
\newblock Generative network complex for the automated generation of drug-like
  molecules.
\newblock {\em Journal of chemical information and modeling},
  60(12):5682--5698, 2020.

\bibitem{gao20202d}
Kaifu Gao, Duc~Duy Nguyen, Vishnu Sresht, Alan~M Mathiowetz, Meihua Tu, and
  Guo-Wei Wei.
\newblock Are 2d fingerprints still valuable for drug discovery?
\newblock {\em Physical Chemistry Chemical Physics}, 22(16):8373--8390, 2020.

\bibitem{li2017deep}
Hang Li.
\newblock Deep learning for natural language processing: advantages and
  challenges.
\newblock {\em National Science Review}, 2017.

\bibitem{yuan2021tokens}
Li~Yuan, Yunpeng Chen, Tao Wang, Weihao Yu, Yujun Shi, Zihang Jiang, Francis~EH
  Tay, Jiashi Feng, and Shuicheng Yan.
\newblock Tokens-to-token vit: Training vision transformers from scratch on
  imagenet.
\newblock {\em arXiv preprint arXiv:2101.11986}, 2021.

\bibitem{singh2021rna}
Jaswinder Singh, Kuldip Paliwal, Jaspreet Singh, and Yaoqi Zhou.
\newblock Rna backbone torsion and pseudotorsion angle prediction using dilated
  convolutional neural networks.
\newblock {\em Journal of Chemical Information and Modeling}, 2021.

\bibitem{devlin2018bert}
Jacob Devlin, Ming-Wei Chang, Kenton Lee, and Kristina Toutanova.
\newblock Bert: Pre-training of deep bidirectional transformers for language
  understanding.
\newblock {\em arXiv preprint arXiv:1810.04805}, 2018.

\bibitem{chen2021algebraic}
Dong Chen, Kaifu Gao, Duc~Duy Nguyen, Xin Chen, Yi~Jiang, Guo-Wei Wei, and Feng
  Pan.
\newblock {Algebraic graph-assisted bidirectional transformers for molecular
  property prediction}.
\newblock {\em Nature Communications}, 12:3521, 2021.

\bibitem{weininger1988smiles}
David Weininger.
\newblock Smiles, a chemical language and information system. 1. introduction
  to methodology and encoding rules.
\newblock {\em Journal of chemical information and computer sciences},
  28(1):31--36, 1988.

\bibitem{wang2019smiles}
Sheng Wang, Yuzhi Guo, Yuhong Wang, Hongmao Sun, and Junzhou Huang.
\newblock Smiles-bert: large scale unsupervised pre-training for molecular
  property prediction.
\newblock In {\em Proceedings of the 10th ACM international conference on
  bioinformatics, computational biology and health informatics}, pages
  429--436, 2019.

\bibitem{bruice2014organic}
Paula~Y Bruice.
\newblock {\em Organic Chemistry: Pearson New International Edition}.
\newblock Pearson Education Limited, 2014.

\bibitem{blondel1996new}
Arnaud Blondel and Martin Karplus.
\newblock New formulation for derivatives of torsion angles and improper
  torsion angles in molecular mechanics: Elimination of singularities.
\newblock {\em Journal of computational chemistry}, 17(9):1132--1141, 1996.

\bibitem{chi2010toxic}
Zhenxing Chi, Rutao Liu, Bingjun Yang, and Hao Zhang.
\newblock Toxic interaction mechanism between oxytetracycline and bovine
  hemoglobin.
\newblock {\em Journal of hazardous materials}, 180(1-3):741--747, 2010.

\bibitem{verma20103d}
Jitender Verma, Vijay~M Khedkar, and Evans~C Coutinho.
\newblock 3d-qsar in drug design-a review.
\newblock {\em Current topics in medicinal chemistry}, 10(1):95--115, 2010.

\bibitem{cang2017topologynet}
Zixuan Cang and Guo-Wei Wei.
\newblock Topologynet: Topology based deep convolutional and multi-task neural
  networks for biomolecular property predictions.
\newblock {\em PLoS computational biology}, 13(7):e1005690, 2017.

\bibitem{meng2020weighted}
Zhenyu Meng, D~Vijay Anand, Yunpeng Lu, Jie Wu, and Kelin Xia.
\newblock Weighted persistent homology for biomolecular data analysis.
\newblock {\em Scientific reports}, 10(1):1--15, 2020.

\bibitem{nguyen2019dg}
Duc~Duy Nguyen and Guo-Wei Wei.
\newblock Dg-gl: Differential geometry-based geometric learning of molecular
  datasets.
\newblock {\em International journal for numerical methods in biomedical
  engineering}, 35(3):e3179, 2019.

\bibitem{nguyen2019agl}
Duc~Duy Nguyen and Guo-Wei Wei.
\newblock Agl-score: Algebraic graph learning score for protein--ligand binding
  scoring, ranking, docking, and screening.
\newblock {\em Journal of chemical information and modeling}, 59(7):3291--3304,
  2019.

\bibitem{zomorodian2005computing}
Afra Zomorodian and Gunnar Carlsson.
\newblock Computing persistent homology.
\newblock {\em Discrete \& Computational Geometry}, 33(2):249--274, 2005.

\bibitem{edelsbrunner2008persistent}
Herbert Edelsbrunner, John Harer, et~al.
\newblock Persistent homology-a survey.
\newblock {\em Contemporary mathematics}, 453:257--282, 2008.

\bibitem{mischaikow2013morse}
Konstantin Mischaikow and Vidit Nanda.
\newblock Morse theory for filtrations and efficient computation of persistent
  homology.
\newblock {\em Discrete \& Computational Geometry}, 50(2):330--353, 2013.

\bibitem{ciocanel2021topological}
Maria-Veronica Ciocanel, Riley Juenemann, Adriana~T Dawes, and Scott~A
  McKinley.
\newblock Topological data analysis approaches to uncovering the timing of ring
  structure onset in filamentous networks.
\newblock {\em Bulletin of Mathematical Biology}, 83(3):1--25, 2021.

\bibitem{wang2020persistent}
Rui Wang, Duc~Duy Nguyen, and Guo-Wei Wei.
\newblock Persistent spectral graph.
\newblock {\em International journal for numerical methods in biomedical
  engineering}, 36(9):e3376, 2020.

\bibitem{jiang2020boosting}
Jian Jiang, Rui Wang, Menglun Wang, Kaifu Gao, Duc~Duy Nguyen, and Guo-Wei Wei.
\newblock Boosting tree-assisted multitask deep learning for small scientific
  datasets.
\newblock {\em Journal of chemical information and modeling}, 60(3):1235--1244,
  2020.

\bibitem{warszycki2021pharmacoprint}
Dawid Warszycki, {\L}ukasz Struski, Marek Śmieja, Rafa{\l} Kafel, and
  Rafa{\l} Kurczab.
\newblock Pharmacoprint: A combination of a pharmacophore fingerprint and
  artificial intelligence as a tool for computer-aided drug design.
\newblock {\em Journal of Chemical Information and Modeling},
  61(10):5054--5065, 2021.

\bibitem{czarnecki2015robust}
Wojciech~M Czarnecki, Sabina Podlewska, and Andrzej~J Bojarski.
\newblock Robust optimization of svm hyperparameters in the classification of
  bioactive compounds.
\newblock {\em Journal of cheminformatics}, 7(1):1--15, 2015.

\bibitem{smusz2015multi}
Sabina Smusz, Stefan Mordalski, Jagna Witek, Krzysztof Rataj, Rafa{\l} Kafel,
  and Andrzej~J Bojarski.
\newblock Multi-step protocol for automatic evaluation of docking results based
  on machine learning methods a case study of serotonin receptors 5-ht6 and
  5-ht7.
\newblock {\em Journal of chemical information and modeling}, 55(4):823--832,
  2015.

\bibitem{lee2021toward}
Kuo~Hao Lee, Andrew~D Fant, Jiqing Guo, Andy Guan, Joslyn Jung, Mary
  Kudaibergenova, Williams~E Miranda, Therese Ku, Jianjing Cao, Soren Wacker,
  et~al.
\newblock Toward reducing herg affinities for dat inhibitors with a combined
  machine learning and molecular modeling approach.
\newblock {\em Journal of Chemical Information and Modeling}, 61(9):4266--4279,
  2021.

\bibitem{Feng2022}
Hongsong Feng, Kaifu Gao, Dong Chen, Li~Shen, Alfred~J. Robison, Edmund
  Ellsworth, and Guo~Wei Wei.
\newblock {Machine Learning Analysis of Cocaine Addiction Informed by DAT,
  SERT, and NET-Based Interactome Networks}.
\newblock {\em Journal of Chemical Theory and Computation}, 18(4):2703--2719,
  2022.

\bibitem{robinson2008opposing}
Emma~SJ Robinson, Jeffrey~W Dalley, David~EH Theobald, Jeffrey~C Glennon,
  Marie~A Pezze, Emily~R Murphy, and Trevor~W Robbins.
\newblock Opposing roles for 5-ht2a and 5-ht2c receptors in the nucleus
  accumbens on inhibitory response control in the 5-choice serial reaction time
  task.
\newblock {\em Neuropsychopharmacology}, 33(10):2398--2406, 2008.

\bibitem{khanna2015buprenorphine}
Ish~K Khanna and Sivaram Pillarisetti.
\newblock Buprenorphine--an attractive opioid with underutilized potential in
  treatment of chronic pain.
\newblock {\em Journal of pain research}, 8:859, 2015.

\bibitem{veilleux2010review}
Jennifer~C Veilleux, Peter~J Colvin, Jennifer Anderson, Catherine York, and
  Adrienne~J Heinz.
\newblock A review of opioid dependence treatment: pharmacological and
  psychosocial interventions to treat opioid addiction.
\newblock {\em Clinical psychology review}, 30(2):155--166, 2010.

\bibitem{carney2020dihydrocodeine}
Tara Carney, Marie~Claire Van~Hout, Ian Norman, Siphokazi Dada, Nandi
  Siegfried, and Charles~DH Parry.
\newblock Dihydrocodeine for detoxification and maintenance treatment in
  individuals with opiate use disorders.
\newblock {\em Cochrane Database of Systematic Reviews}, (2), 2020.

\bibitem{raynor1994pharmacological}
Karen Raynor, Haeyoung Kong, Yan Chen, KAZUKI Yasuda, Lei Yu, Graeme~I Bell,
  and TERRY Reisine.
\newblock Pharmacological characterization of the cloned kappa-, delta-, and
  mu-opioid receptors.
\newblock {\em Molecular pharmacology}, 45(2):330--334, 1994.

\bibitem{newman2021new}
Amy~Hauck Newman, Therese Ku, Chloe~J Jordan, Alessandro Bonifazi, and
  Zheng-Xiong Xi.
\newblock New drugs, old targets: tweaking the dopamine system to treat
  psychostimulant use disorders.
\newblock {\em Annu. Rev. Pharmacol. Toxicol}, 61:609--628, 2021.

\bibitem{efange1998modified}
Simon~MN Efange, Deborah~C Mash, Anil~B Khare, and Quinjie Ouyang.
\newblock Modified ibogaine fragments: Synthesis and preliminary
  pharmacological characterization of 3-ethyl-5-phenyl-1, 2, 3, 4, 5,
  6-hexahydroazepino [4, 5-b] benzothiophenes.
\newblock {\em Journal of medicinal chemistry}, 41(23):4486--4491, 1998.

\bibitem{koenig2015anti}
Xaver Koenig and Karlheinz Hilber.
\newblock The anti-addiction drug ibogaine and the heart: a delicate relation.
\newblock {\em Molecules}, 20(2):2208--2228, 2015.

\bibitem{hozumi2022ccp}
Yuta Hozumi, Rui Wang, and Guo-Wei Wei.
\newblock Ccp: Correlated clustering and projection for dimensionality
  reduction.
\newblock {\em arXiv preprint arXiv:2206.04189}, 2022.

\bibitem{https://doi.org/10.48550/arxiv.1802.03426}
Leland McInnes, John Healy, and James Melville.
\newblock Umap: Uniform manifold approximation and projection for dimension
  reduction, 2018.

\bibitem{gaulton2012chembl}
Anna Gaulton, Louisa~J Bellis, A~Patricia Bento, Jon Chambers, Mark Davies,
  Anne Hersey, Yvonne Light, Shaun McGlinchey, David Michalovich, Bissan
  Al-Lazikani, et~al.
\newblock Chembl: a large-scale bioactivity database for drug discovery.
\newblock {\em Nucleic acids research}, 40(D1):D1100--D1107, 2012.

\bibitem{smieja2016average}
Marek {\'S}mieja and Dawid Warszycki.
\newblock Average information content maximization—a new approach for
  fingerprint hybridization and reduction.
\newblock {\em PloS one}, 11(1):e0146666, 2016.

\bibitem{zhao2020rham}
Rundong Zhao, Menglun Wang, Jiahui Chen, Yiying Tong, and Guo-Wei Wei.
\newblock The de rham--hodge analysis and modeling of biomolecules.
\newblock {\em Bulletin of mathematical biology}, 82(8):1--38, 2020.

\bibitem{chen2021evolutionary}
Jiahui Chen, Rundong Zhao, Yiying Tong, and Guo-Wei Wei.
\newblock Evolutionary de rham-hodge method.
\newblock {\em Discrete and continuous dynamical systems. Series B},
  26(7):3785, 2021.

\bibitem{kamber1987rham}
Franz~W Kamber and Philippe Tondeur.
\newblock de rham-hodge theory for riemannian foliations.
\newblock {\em Mathematische Annalen}, 277(3):415--431, 1987.

\bibitem{vaswani2017attention}
Ashish Vaswani, Noam Shazeer, Niki Parmar, Jakob Uszkoreit, Llion Jones,
  Aidan~N Gomez, {\L}ukasz Kaiser, and Illia Polosukhin.
\newblock Attention is all you need.
\newblock In {\em Advances in neural information processing systems}, pages
  5998--6008, 2017.

\bibitem{chen2021extracting}
Dong Chen, Jiaxin Zheng, Guo-Wei Wei, and Feng Pan.
\newblock Extracting predictive representations from hundreds of millions of
  molecules.
\newblock {\em The Journal of Physical Chemistry Letters}, 12(44):10793--10801,
  2021.

\end{thebibliography}
\end{document}